\documentclass[twocolumn]{aastex63}
\usepackage{amsmath}
\usepackage{comment}

\received{}
\revised{\today}
\accepted{}

\submitjournal{ApJ}

\begin{document}

\title{
A FORECAST OF THE SENSITIVITY ON THE MEASUREMENT 
OF THE OPTICAL DEPTH TO REIONIZATION
WITH THE GROUNDBIRD EXPERIMENT
}


\correspondingauthor{Eunil Won}
\email{eunilwon@korea.ac.kr}

\author[0000-0002-2247-3728]{K. Lee}
\affiliation{Department of Physics, Korea University, Seoul, 02841, Republic of Korea}


\author[0000-0001-5479-0034]{R. T. G\'enova-Santos}
\affiliation{Instituto de Astrof\'isica de Canarias, E38205 La Laguna, Tenerife, Canary Islands, Spain}
\affiliation{Departamento de Astrof\'isica, Universidad de La Laguna, E38206 La Laguna, Tenerife, Canary Islands, Spain}

\author[0000-0001-6830-8309]{M. Hazumi}
\affiliation{High Energy Accelerator Research Organization (KEK), Tsukuba, Ibaraki, 305-0801, Japan}
\affiliation{Japan Aerospace Exploration Agency (JAXA), Institute of Space and Astronautical Science (ISAS), Sagamihara, Kanagawa 252-5210, Japan}
\affiliation{Kavli Institute for the Physics and Mathematics of the Universe (Kavli IPMU, WPI), UTIAS, The University of Tokyo, Kashiwa, Chiba 277-8583, Japan}
\affiliation{The Graduate University for Advanced Studies (SOKENDAI), Miura District, Kanagawa 240-0115, Hayama, Japan}

\author[0000-0002-0403-3729]{S. Honda}
\affiliation{Physics Department, Kyoto University, Kyoto, 606-8502, Japan}


\author{H. Kutsuma}
\affiliation{Department of Physics, Tohoku University, 6-3 Aramaki-Aoba, Aoba-ku, Sendai, Miyagi 980-8578, Japan}
\affiliation{The Institute of Physical and Chemical Research (RIKEN), 519-1399 Aramaki-Aoba, Aoba-ku, Sendai, Miyagi 980-0845, Japan}


\author[0000-0002-5902-2672]{S. Oguri}
\affiliation{Japan Aerospace Exploration Agency (JAXA), Institute of Space and Astronautical Science (ISAS), Sagamihara, Kanagawa 252-5210, Japan}

\author{C. Otani}
\affiliation{Department of Physics, Tohoku University, 6-3 Aramaki-Aoba, Aoba-ku, Sendai, Miyagi 980-8578, Japan}
\affiliation{The Institute of Physical and Chemical Research (RIKEN), 519-1399 Aramaki-Aoba, Aoba-ku, Sendai, Miyagi 980-0845, Japan}

\author[0000-0003-3412-2586]{M. W. Peel}
\affiliation{Instituto de Astrof\'isica de Canarias, E38205 - La Laguna, Tenerife, Canary Islands, Spain}
\affiliation{Departamento de Astrof\'isica, Universidad de La Laguna, E38206 La Laguna, Tenerife, Canary Islands, Spain}




\author{Y. Sueno}
\affiliation{Physics Department, Kyoto University, Kyoto, 606-8502, Japan}

\author[0000-0001-6816-8123]{J. Suzuki}
\affiliation{Physics Department, Kyoto University, Kyoto, 606-8502, Japan}

\author{O. Tajima}
\affiliation{Physics Department, Kyoto University, Kyoto, 606-8502, Japan}

\author[0000-0002-4245-7442]{E. Won}
\affiliation{Department of Physics, Korea University, Seoul, 02841, Republic of Korea}


\begin{abstract}
We compute the expected sensitivity on
measurements of optical depth to reionization 
for a ground-based experiment at Teide Observatory. 
We simulate polarized partial sky maps 
for the GroundBIRD experiment at the frequencies 
145 and 220 GHz. We perform fits for 
the simulated maps with our pixel-based likelihood 
to extract the optical depth to reionization. 
The noise levels of polarization maps are estimated as 
110 $\mu\mathrm{K~arcmin}$ and 780 $ \mu\mathrm{K~arcmin}$
for 145 and 220 GHz, respectively,
by assuming a three-year 
observing campaign and sky coverages of 
0.537 for 145 GHz and 0.462 for 220 GHz.
Our sensitivities for the optical depth to reionization are 
found to be $\sigma_\tau$=0.030 with the simulated 
GroundBIRD maps, and $\sigma_\tau$=0.012 by combining with 
the simulated QUIJOTE maps at 11, 13, 17, 19, 30, and 40 GHz.
\end{abstract}



\section{Introduction} 
\label{sec:intro} 

A model of rapid accelerated expansion of the early universe \citep{Starobinsky1980, Guth1981,  Sato1981, Albrecht1982, Linde1982}, required 
in order to resolve the horizon and flatness problems of the standard
model of cosmology, predicts the production of two distinct polarized patterns
in the cosmic microwave background (CMB) radiation due to density
fluctuation and the production of primordial gravitational waves \citep{Kamionkowski1997,Seljak1997,Zaldarriaga1997,Kamionkowski1997_2}. 
Along with this possible inflationary expansion scenario of
the early universe, the standard model of  cosmology, or a spatially-flat six-parameter model of the universe ($\Lambda$CDM),
having a cosmological constant and dark matter as dominant forms of the energy density in
the universe at present, agrees 
remarkably well with observational data to date. 
In both cases, the CMB polarization plays a key role 
in discovering the physics of the early universe. 
Among six parameters in the $\Lambda$CDM model of the standard cosmology, 
the optical depth to reionization ($\tau$), 
a dimensionless quantity that provides a measure 
of the line-of-sight free-electron opacity to CMB radiation, 
has important cosmological implications.

Of the six parameters in the $\Lambda$CDM model mentioned, 
$\tau$ is the only one that is measured 
with sensitivity worse than 1\% at present,
$\tau = 0.054 \pm 0.007$ \citep{planck2018vi}.
This measurement also constrains the sum of the neutrino masses, 
one of the critical parameters 
in describing the evolution of the early universe.
We can determine the sum of the neutrino masses by comparing the 
primordial amplitude of CMB, $A_s$, and the amplitude of CMB at 
small scales. It gives, however, the combination of $A_se^{-\tau}$ only, 
and thus a precise measurement of $\tau$ is required to better constrain
the neutrino mass via $A_s$ \citep{Allison2015, Bryan2018, Ferraro2018, planck2018vi}.

At large angular scales in the polarization power spectrum of the CMB 
($\ell \lesssim 10$ where $\ell$ is the multipole index 
in the angular power spectrum), 
anisotropies are created by the rescattering 
of the local temperature quadrupole  \citep{Zaldarriaga1997a, planck2018vi}.
These lead to a ``bump'' today in the large-scale polarization 
power spectrum at the Hubble scale during reionization \citep{Hu1997}.
The amplitude of the bump scales like a power law of
$\tau^2$ 
and $\tau$ is therefore mostly constrained
by the large-scale polarization measurements 
from the Planck satellite \citep{planck2018vi}.
Recent space-based measurements of $\tau$ have 
a clear systematic tendency for the central value to decrease, 
and the large angular scale polarization measurements become crucial.
In the near future, we expect a part of
recently-started ground-based experiments such as 
ACTpol/SPTpol \citep{ACTpol, SPTpol}, 
BICEP/Keck Array \citep{BICEP}, 
CLASS \citep{Watts2018},
GroundBIRD \citep{Tajima2012gb, won2018gb}, 
POLARBEAR \citep{Polarbear, Kermish:2012eh},
Simons Array \citep{simons_array}, 
Simons Observatory \citep{simons_observatory}, 
STRIP \citep{STRIP}, 
and QUIJOTE \citep{quijote2015}
to provide such measurements.
Among these, CLASS, GroundBIRD, and QUIJOTE are sensitive
to the polarization of CMB at large angular scales ($\ell \lesssim 20$).

In this paper, we calculate the expected sensitivity 
of the GroundBIRD experiment to $\tau$, 
based on simulated maps with the GroundBIRD configuration. 
A joint measurement with QUIJOTE, which has six frequency bands 
between 11 and 40 GHz, 
will provide better systematic error control for the foregrounds. 
We also estimate the sensitivity of this combined data analysis.

In section \ref{sec:telescopes}, we describe the GroundBIRD  and QUIJOTE experiments;
in section \ref{sec:sim}, we describe our process to prepare the data; 
in section \ref{sec:foregroundcleaning}, we describe the foreground cleaning;
in section \ref{sec:likelihood}, we describe our pixel-likelihood method;
in section \ref{sec:result}, we discuss the result of the analysis; and 
in section \ref{sec:conclusions}, we present our conclusions.

\section{The GroundBIRD and QUIJOTE telescopes}
\label{sec:telescopes}

GroundBIRD 
(Ground-based Background Image Radiation Detector)
is a ground-based experiment that 
measures the polarization of the CMB radiation. 
It will cover about 45\% of the sky 
from the Teide Observatory
($28 \arcdeg 18\arcmin~\mathrm{N}$ and $16 \arcdeg 30 \arcmin~\mathrm{W}$ in the northern hemisphere, 
2400 m above mean sea level)
in Tenerife, at two frequencies (145 and 220 GHz).
This large sky coverage is obtained by an off-axis rotation of the telescope, 
as fast as 20 revolutions per minute (rpm), 
with the aim of minimizing impact from 
the instrument and atmospheric $1/f$ noise.
Here, we refer to $f$ as the frequency 
in the power spectral density of the time-ordered data (TOD). 
The GroundBIRD telescope also 
has cryogenic mirrors in a crossed-Dragone optical set-up
at a temperature of 4 K to minimize unwanted radiation 
from the optics reaching the photon sensors \citep{Tajima2012gb}. 
GroundBIRD's photon sensors are kinetic inductance detectors (KIDs) \citep{Day2003}, 
and we plan to use six modules with 138 and 
one module with 23 polarization-sensitive detectors for 145 and 220 GHz, respectively.
The GroundBIRD telescope will be a testbed for high-speed
rotation scanning with a cryogenic mirror \citep{fastrotationcooling}
and polarization measurements of CMB with KID \citep{nika2}.

The QUIJOTE (Q-U-I JOint Tenerife) is a polarization experiment \citep{Alberto_2012},
also located at Teide Observatory, consisting of two 2.3-m Cross-Dragone telescopes
which are equipped with three radiometer-based instruments: the MFI (multifrequency
instrument, 11, 13, 17 and 19 GHz), the TGI (thirty-gigahertz instrument, 30 GHz), and the FGI
(forty-gigahertz instrument, 40 GHz). By observing at elevations down to 30 deg, QUIJOTE
achieves a sky coverage of up to $f_\mathrm{sky} \sim 70\%$. The combination of large-angular scale and
low-frequency spectral coverage in multiple frequency bands makes it a unique experiment,
especially with regard to its capability to provide accurate information about the spectrum
and spatial distribution of the synchrotron emission. 

The combination of QUIJOTE and GroundBIRD, both observing a similar sky footprint from 
the same site and using similar scanning strategies, provides a powerful way to perform 
a joint characterization of both foreground contaminants.

\section{Generation and reconstruction of maps}
\label{sec:sim}

In the experiment, 
the TOD correlates with the intensities of the sky signal given the direction of the telescope 
and the polarization antenna directions. With the TOD, we construct the sky maps for different
frequencies through a map-making process. 
For our simulation study, however, we first simulate the template sky maps of different 
frequencies. Our TOD is then obtained by simulating observations with these template sky maps.
We obtain our final maps through map-making from further procedures described below. 

    \begin{figure*}    
        \gridline{
            \fig{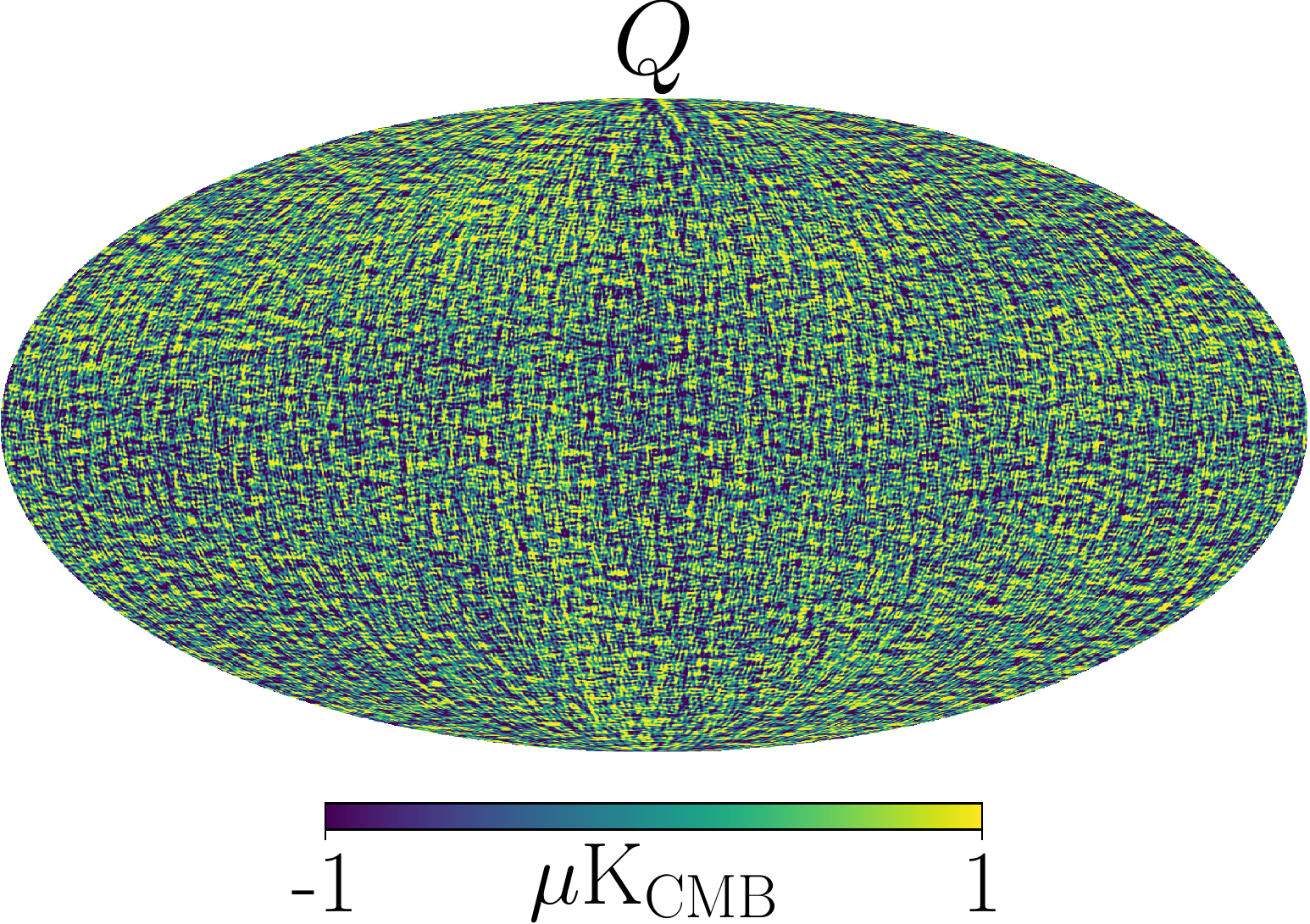}{0.35\textwidth}{(a) Simulated CMB $Q$}
            \fig{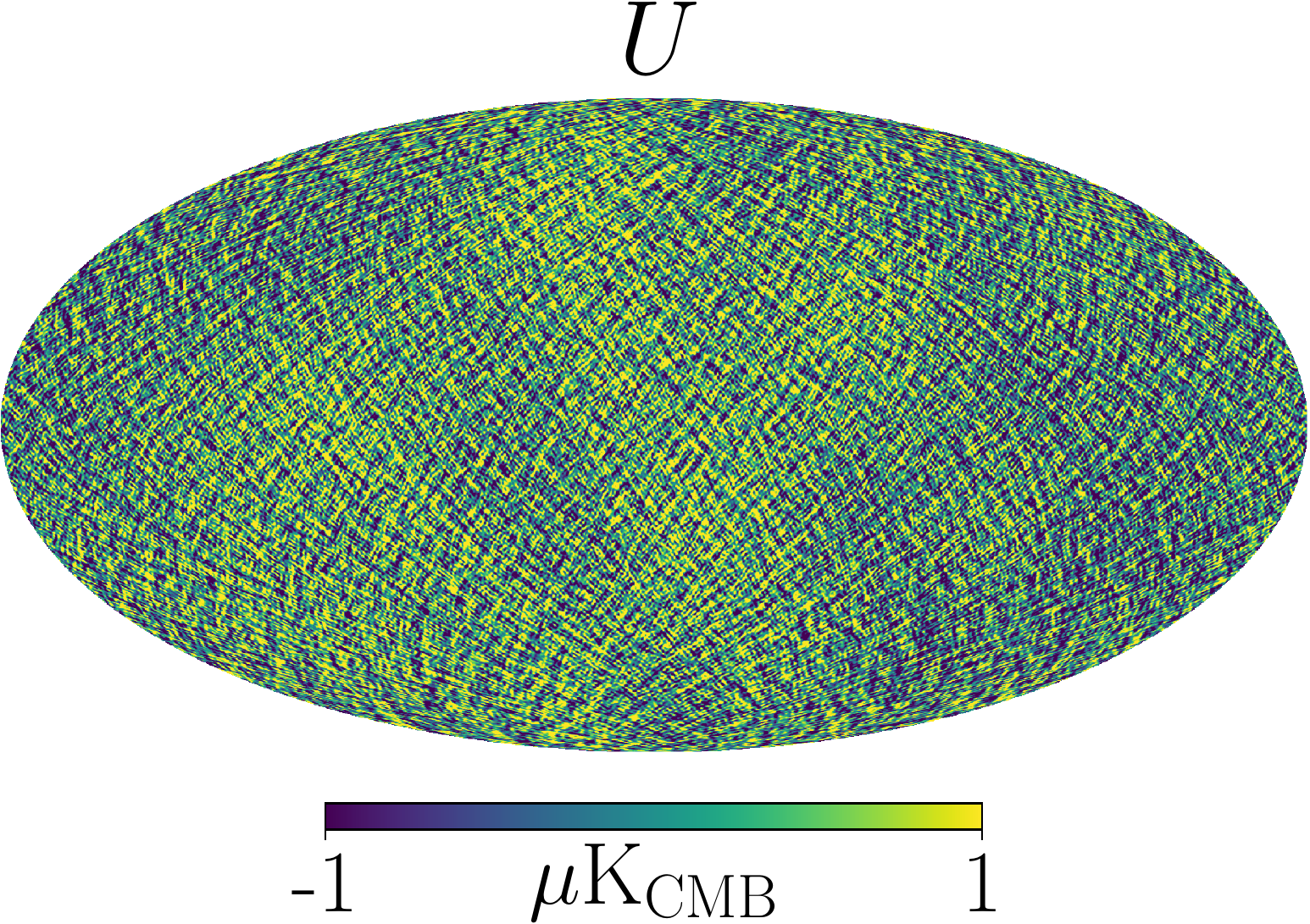}{0.35\textwidth}{(b) Simulated CMB $U$}
        }
        \caption{ 
            Simulated CMB polarization maps are shown: (a) and (b) show 
            the $Q$ and $U$ maps, respectively. 
            The temperature unit, $\mu \mathrm{K_{CMB}}$, 
            is the CMB temperature or thermodynamic temperature. 
            These maps are used as the template for CMB TOD. 
            Here $N_\textrm{side} = 1024$ and the Galactic coordinate system are used. 
            All the maps are smoothed with a beamwidth of $0.6 \arcdeg$. 
        } 
        \label{fig:simulated_cmb}
    \end{figure*}

\subsection{CMB}
   
We use the {\tt CAMB} \citep{Lewis:1999bs}\footnote{\url{https://camb.info/}} software 
for generating the theoretical CMB angular power spectra. 
All the cosmological parameters are taken 
from the latest Planck result \citep{planck2018vi} unless otherwise stated.
For $B$-mode spectrum, 
primordial gravitational waves are excluded
but the lensing effect is included.
The maps are synthesized using {\tt SYNFAST} 
in {\tt HEALPix} \citep{healpix}\footnote{\url{http://healpix.sf.net}} software 
with the partition parameter $N_\textrm{side} =1024$ in {\tt HEALPix},
which corresponds to an angular resolution of 3.4 arcmin. 
Figure \ref{fig:simulated_cmb} shows the synthesized CMB polarization maps. 
The Stokes parameters of linear polarization are 
shown as $Q$ and $U$. 
The input spectra for the map synthesis are generated with $\tau=0.05$. 
The synthesized maps are convolved 
with a Gaussian beam with $0.6\arcdeg$ FWHM (full width at half maximum), 
which is the designed beamwidth of the 145 GHz detectors of the GroundBIRD telescope. 
The beamwidth of the 145 GHz detectors is wider than that of the 220 GHz detectors, 
so we choose to smooth both maps with the beamwidth of $0.6\arcdeg$ FWHM 
to have a common resolution.

\subsection{Foregrounds}

    \begin{figure*}  
        \gridline{
            \fig{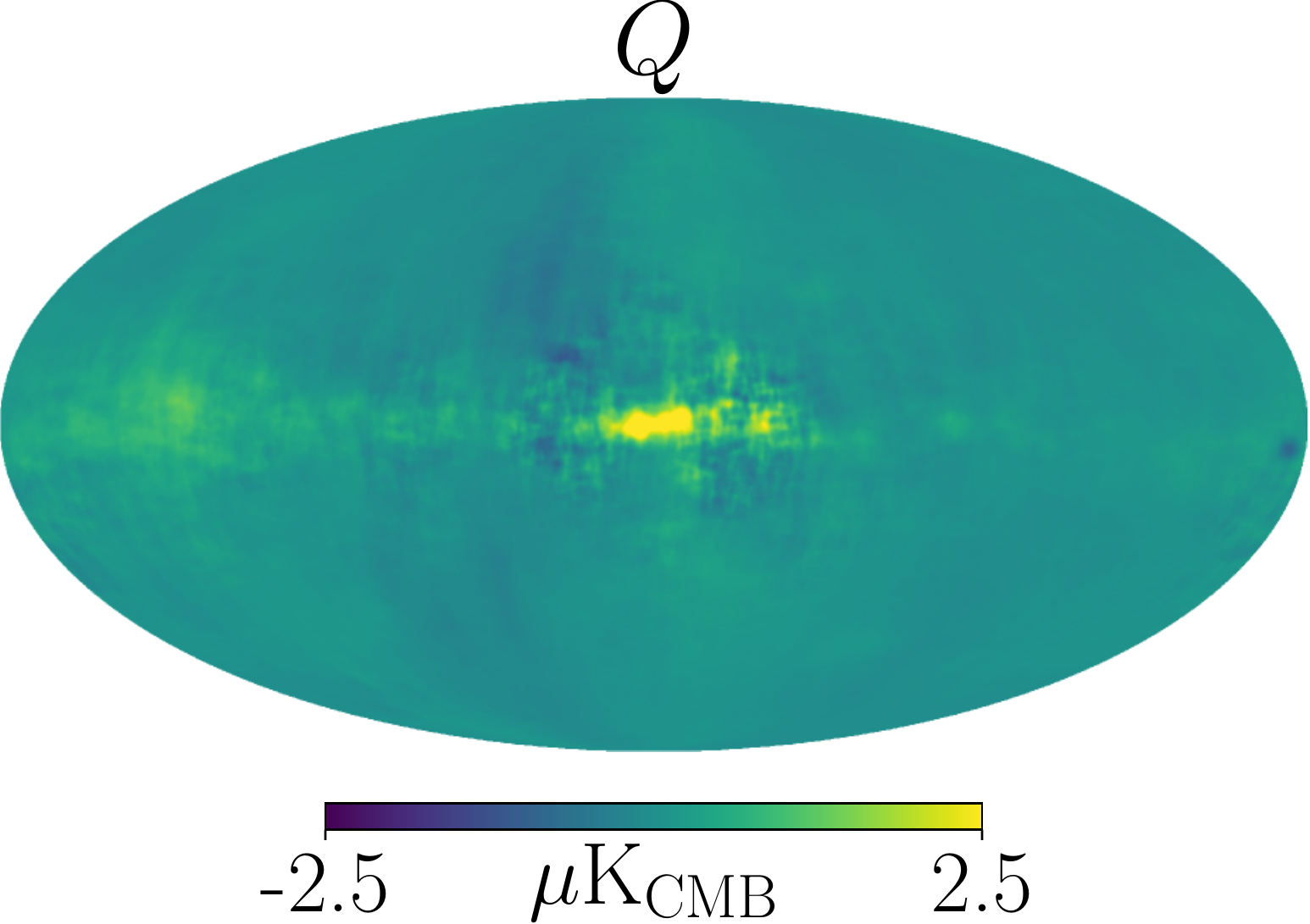}{0.35\textwidth}
            {(a) Synchrotron at 145 GHz, $Q$}
            \fig{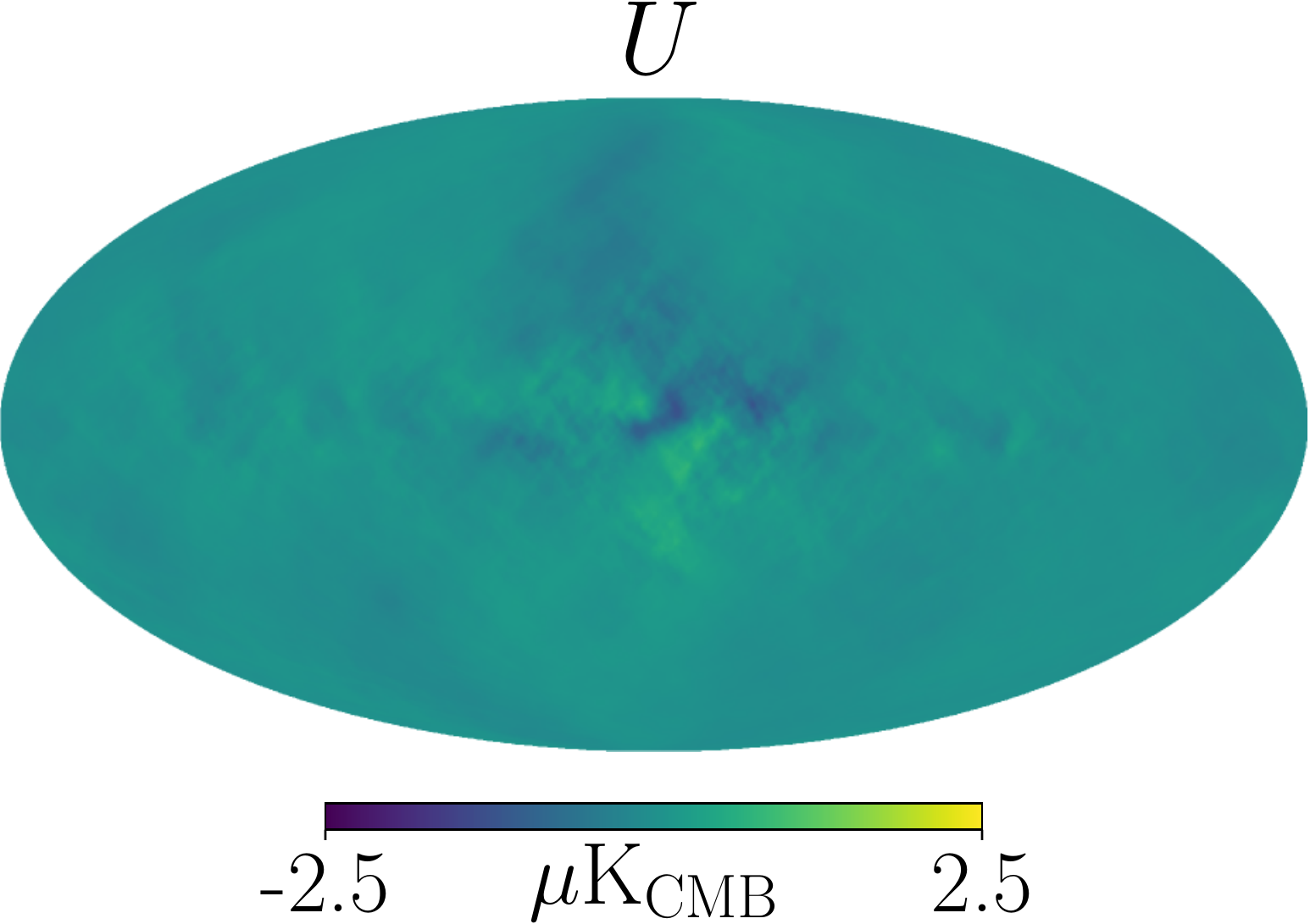}{0.35\textwidth}
            {(b) Synchrotron at 145 GHz, $U$}
        }
        \gridline{
            \fig{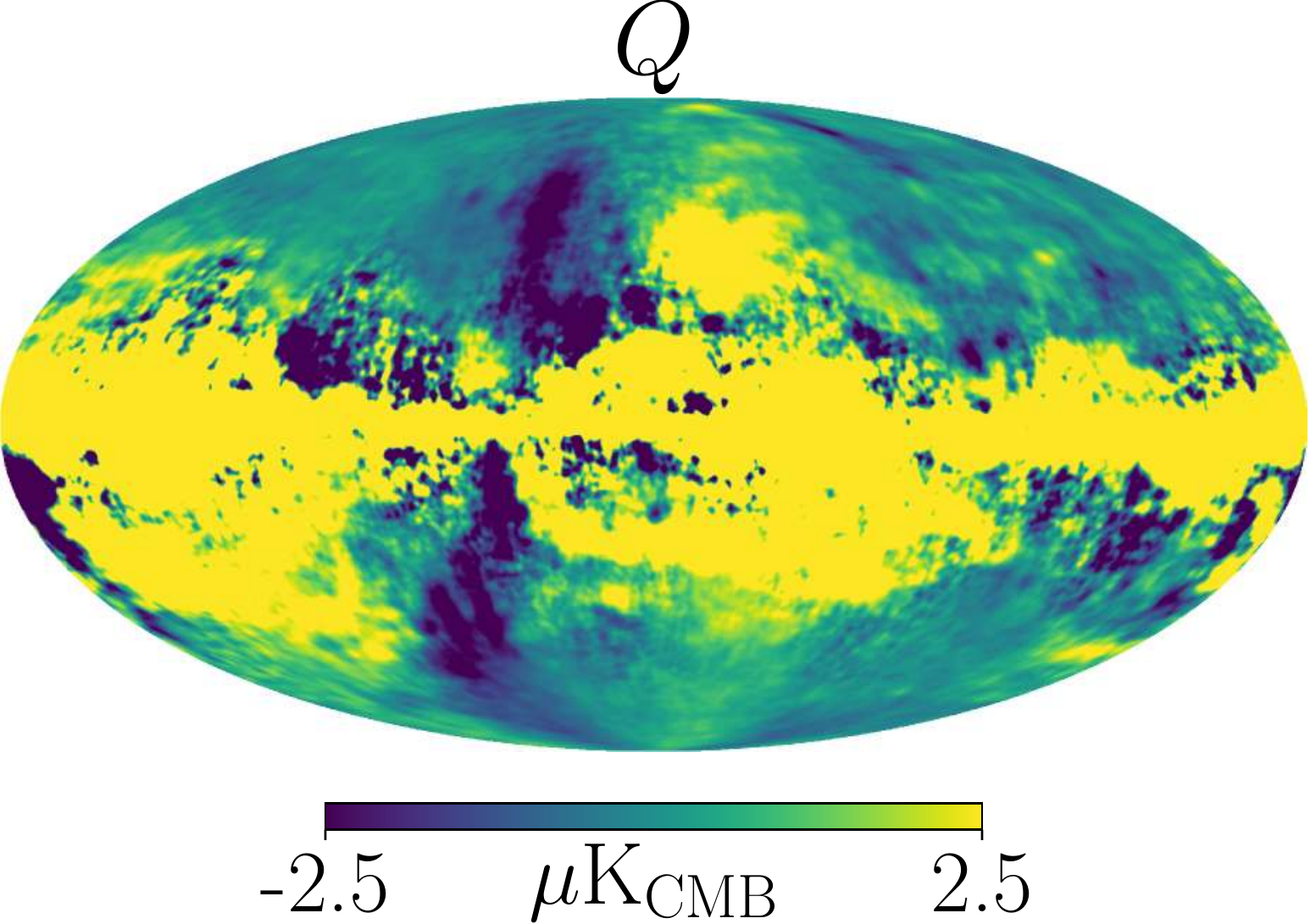}{0.35\textwidth}
            {(c) Dust at 145 GHz, $Q$}
            \fig{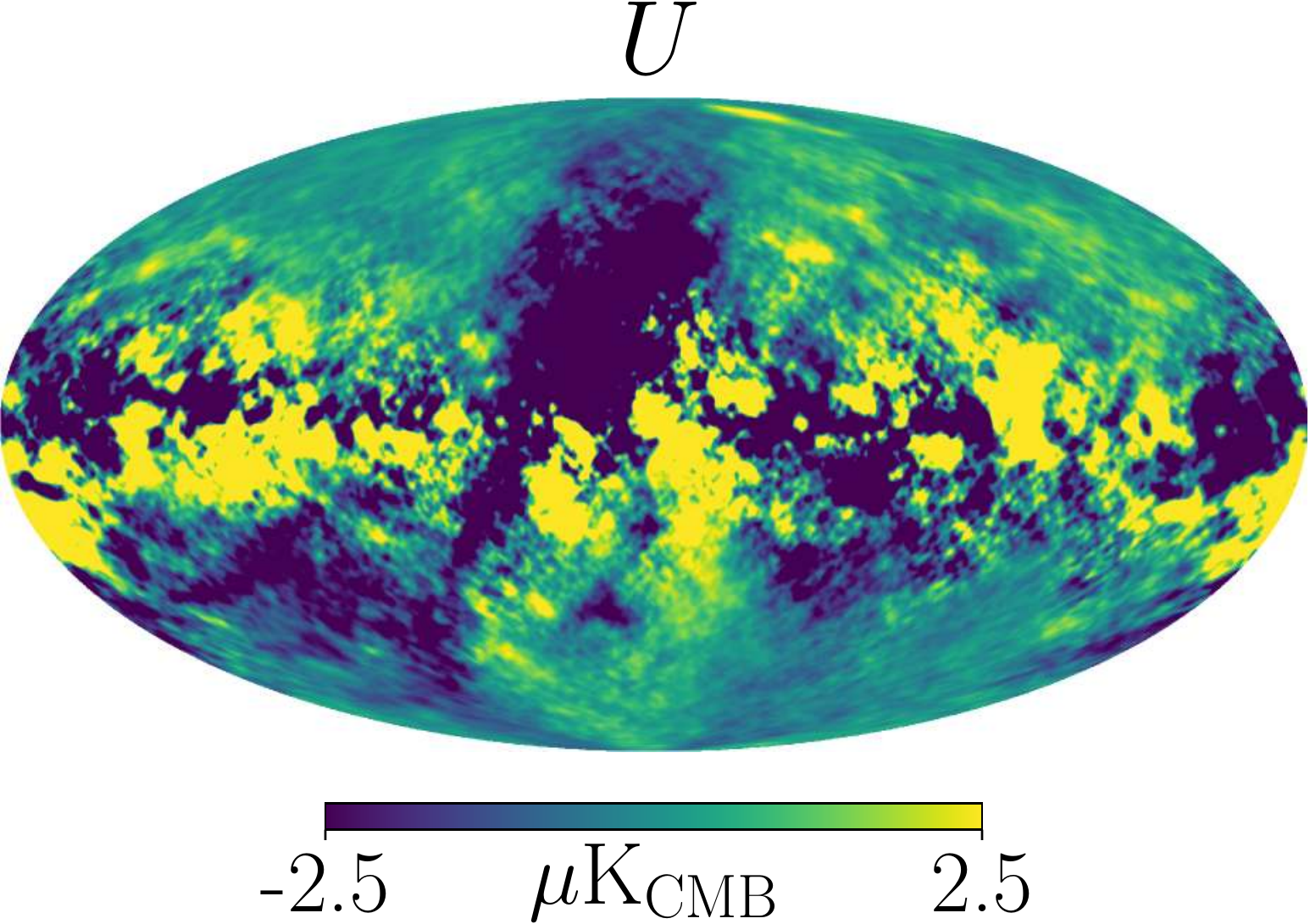}{0.35\textwidth}
            {(d) Dust at 145 GHz, $U$}
        }
        \gridline{
            \fig{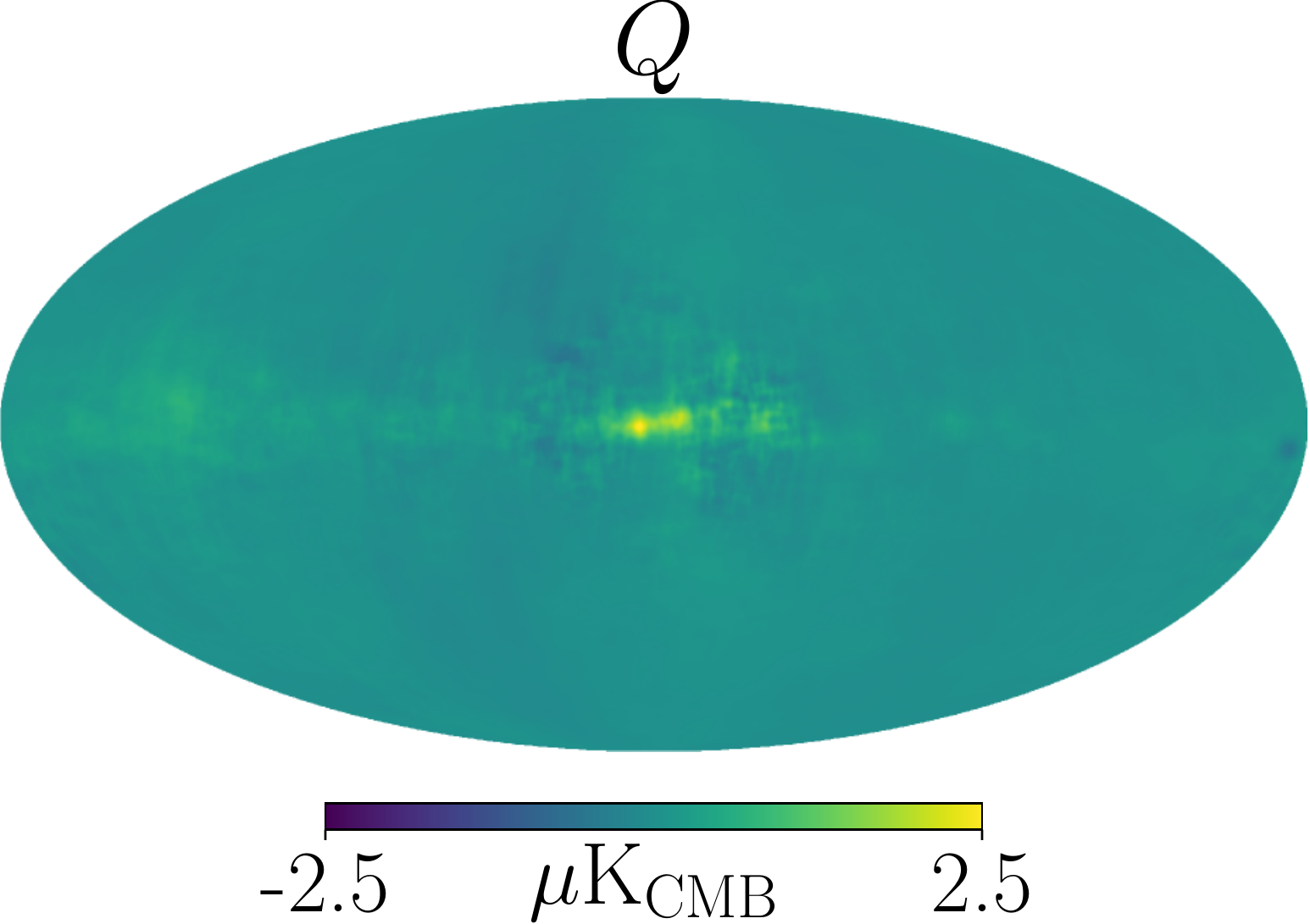}{0.35\textwidth}
            {(e) Synchrotron at 220 GHz, $Q$}
            \fig{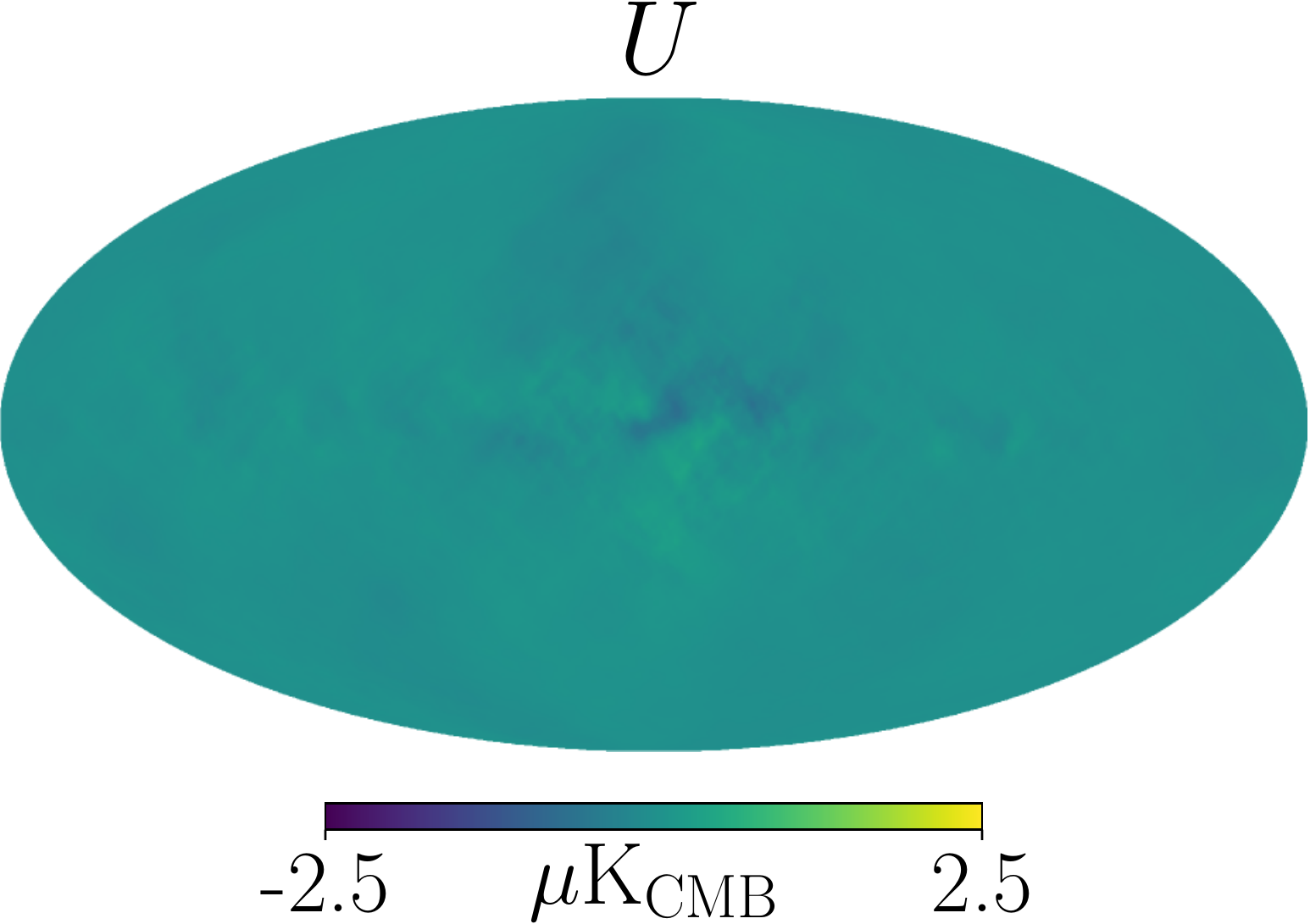}{0.35\textwidth}
            {(f) Synchrotron at 220 GHz, $U$}
        }
        \gridline{
            \fig{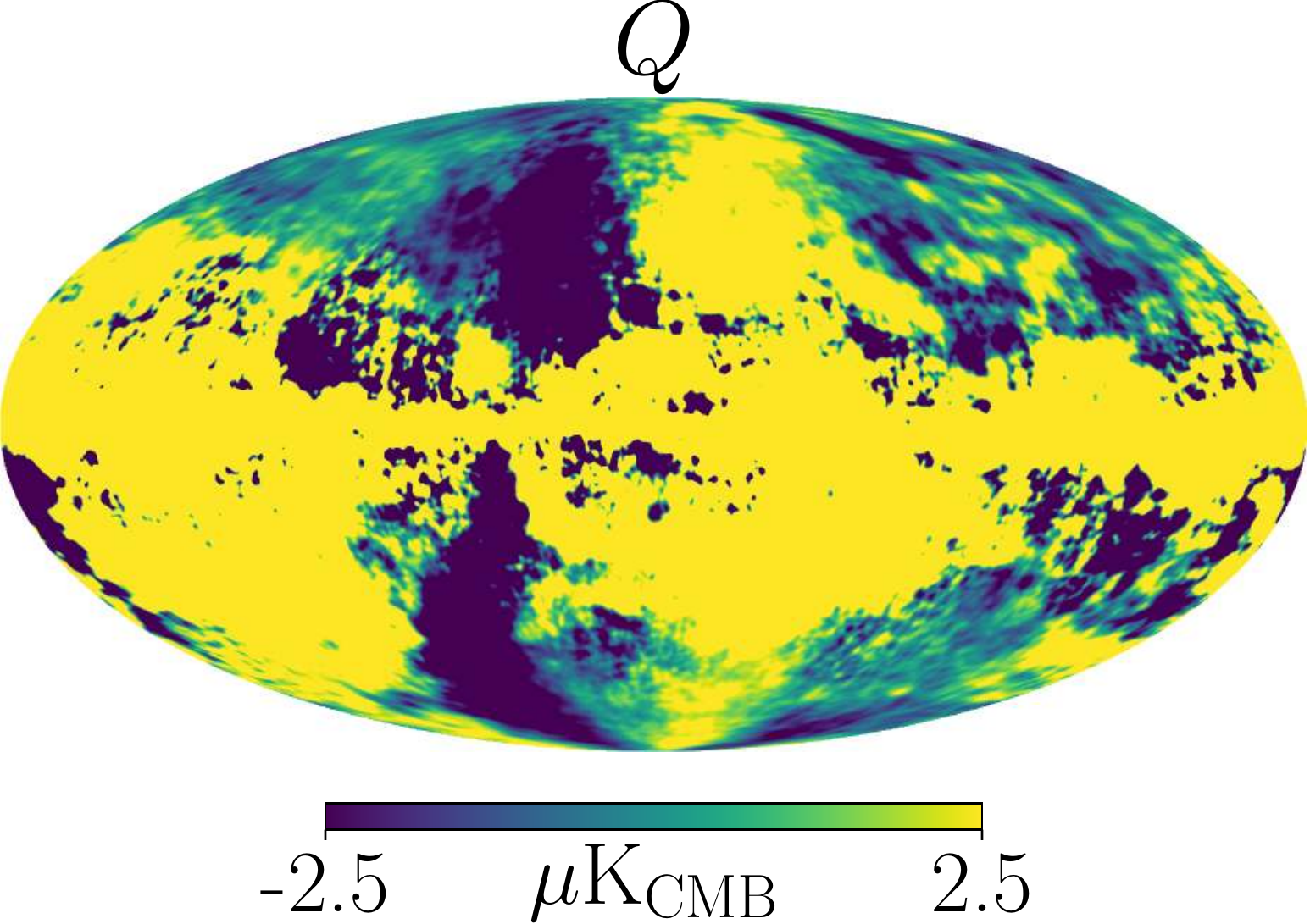}{0.35\textwidth}
            {(g) Dust at 220 GHz, $Q$}
            \fig{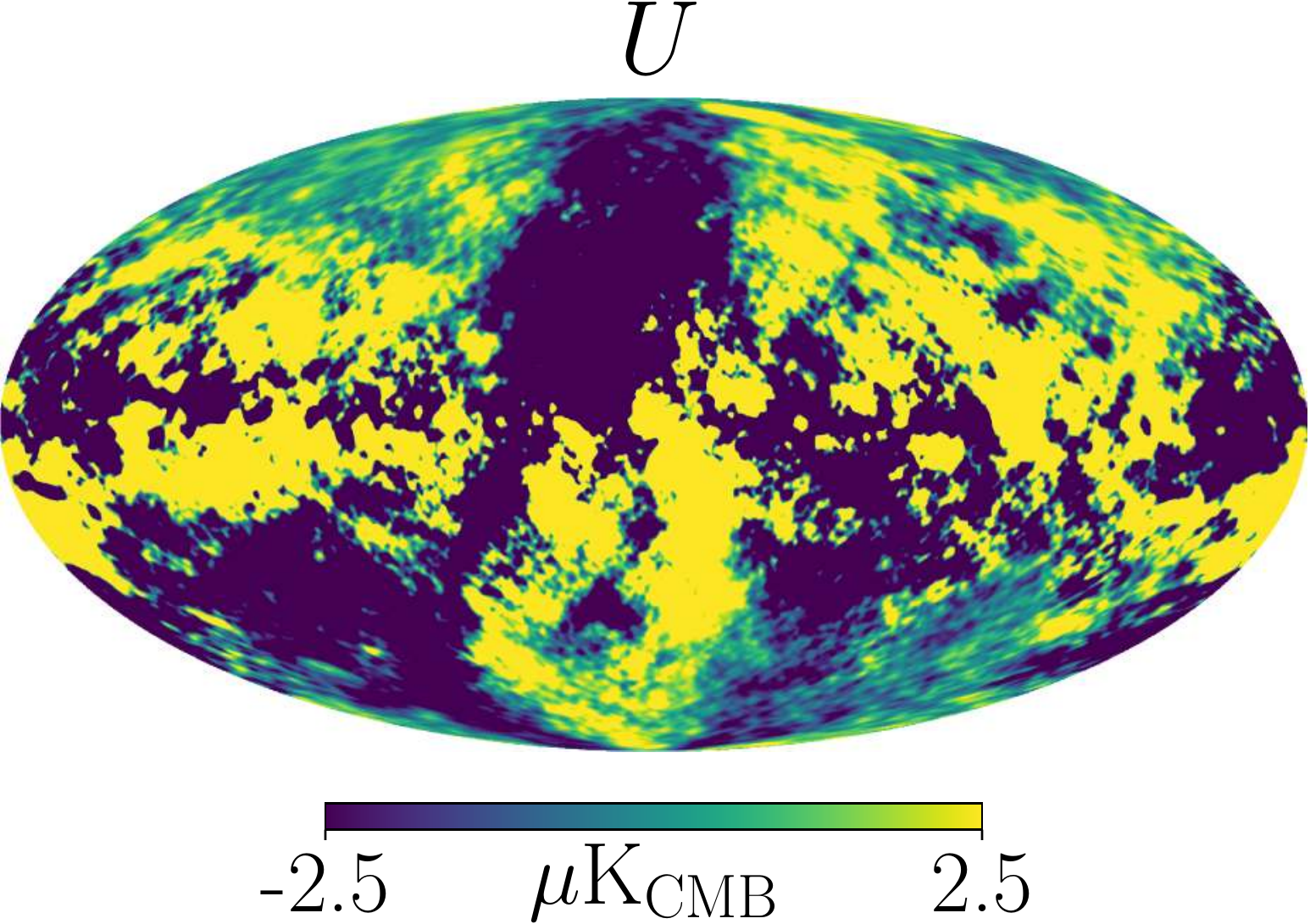}{0.35\textwidth}
            {(h) Dust at 220 GHz, $U$}
        }
        \caption{ 
            A set of simulated foreground maps is shown. 
            The $Q$ and $U$ polarization maps 
            for synchrotron at 145 GHz, dust at 145 GHz, 
            synchrotron at 220 GHz, and dust at 220 GHz are shown. 
            Here $N_\textrm{side} = 1024$, and the Galactic coordinate system is used. 
            The units are $\mu\mathrm{K_{CMB}}$.
            All the maps are smoothed with a beamwidth of $0.6 \arcdeg$, as for CMB maps.
        }
        \label{fig:simulated_foregrounds}
    \end{figure*}
    
The foregrounds are generated 
with the public version of the Python Sky Model 
({\tt PySM}) \citep{PySM}.\footnote{\url{https://github.com/bthorne93/PySM_public}}
The polarization maps include only the synchrotron and dust foregrounds. 
Other foreground components such as free--free or 
anomalous microwave emission components are mostly unpolarized \citep{Ricardo_2015}. 
The WMAP 23 GHz and the Planck 353 GHz polarization maps
are used as the synchrotron and dust templates \citep{PySM}.
The foreground maps for a given frequency are obtained by extrapolating these template maps.
For both of the foreground components,
spatially-varying spectral indices
are used to scale the templates of the components \citep{PySM}.
The foreground maps obtained for GroundBIRD frequencies
are shown in Fig. \ref{fig:simulated_foregrounds}
for each component for each frequency band.
Here $N_\textrm{side}=1024$, and the Galactic coordinate system is used.
Note that the temperature units are the CMB temperatures, $\mu\mathrm{K_{CMB}}$, for all the maps.
As shown in Fig. \ref{fig:simulated_foregrounds}, the dust component is dominant for GroundBIRD frequencies. 

In order to improve the foreground removal further,
we also simulate other frequency maps, for the QUIJOTE frequencies 11, 13, 17, 19, 30, and 40 GHz.
These will be described in the foreground cleaning stage.

\subsection{TOD generation}

We simulate the TOD based on the observation method of the GroundBIRD telescope, 
which scans the sky with a continuous azimuthal rotation at a fixed elevation.
The generated CMB and foreground maps are used for the input to the TOD simulation.
The line of sight and the polarization direction of each detector
are computed by accounting for the rotations of the telescope and the Earth.

We construct two rotational matrices to compute the observation points in celestial coordinates. 
One matrix accounts for the boresight angle, 
elevation, and azimuthal rotation of the telescope. 
The other one accounts for the location of the observing site and the rotation of the Earth. 
As a net result, we obtain the observation points 
in the equatorial coordinate system. 
For each step of the computation, we use the routines in 
{\tt healpy} \citep{healpy} and {\tt astropy} \citep{astropy:2013, astropy:2018}.

We set the elevation of the telescope at $70 \arcdeg$,
and the rotation speed at 20 rpm with a data sampling rate of 1000 samples per second.
The angular distance between two consecutive samples is 2.4 arcmin on the sky, 
which is smaller than the pixel size of the template maps.
It does not cause any sizable effect
because our beamwidth is much larger than the pixel size. 
The TOD is expressed in $\mu\mathrm{K_{CMB}}$.

\subsection{Noise}

The noise in TOD contains white noise and $1/f$ noise.
The white noise is a frequency-independent term due to random fluctuation. 
The $1/f$ noise, however, is a frequency-dependent term that is higher at low frequencies.
It contains any low-frequency drifts due to detector condition changes or 
fluctuation of atmospheric radiation.
For this study, we assume the knee frequency, 
the frequency where the $1/f$ noise is equal to the white noise, 
to be 0.1 Hz. 
Note that we also assume that the exponent of the $1/f$ noise component is equal to unity.
Our estimation for noise equivalent temperature (NET) values of 
the detector and the atmospheric noise combined for GroundBIRD are 
820 $\mu\mathrm{K\sqrt{s}}$ for 145 GHz 
and 2600 $\mu\mathrm{K\sqrt{s}}$ for 220 GHz 
in $\mu\mathrm{K_{CMB}}$. 
The calculation details are given in Appendix \ref{appendix:NET}.
We assume the three-year observation time and $70\%$ observation efficiency.
The numbers of detectors are 138 for 145 GHz and 23 for 220 GHz.
We also assume the sky coverages to be $0.537$ and $0.462$ for 145 and 220 GHz, respectively.
These provide the mean observation times per detector
to be 0.83 $\mathrm{s/arcmin^2}$ for 145 GHz 
and 0.97 $\mathrm{s/arcmin^2}$ for 220 GHz.
These assumptions lead to expected pixel noise levels for $Q$ and $U$ \citep{Wu_2014} of 
110 $\mu\mathrm{K~arcmin}$ for 145 GHz and 
780 $\mu\mathrm{K~arcmin}$ for 220 GHz.
We use these pixel noise levels to generate uniform white noise maps with {\tt SYNFAST}. 
We can also make noise maps by map-making from noise TOD.
For this study, we choose the former method 
to generate multiple noise realizations at a low cost. 

\subsection{Map-making}

    \begin{figure*} 
        \gridline{
            \fig{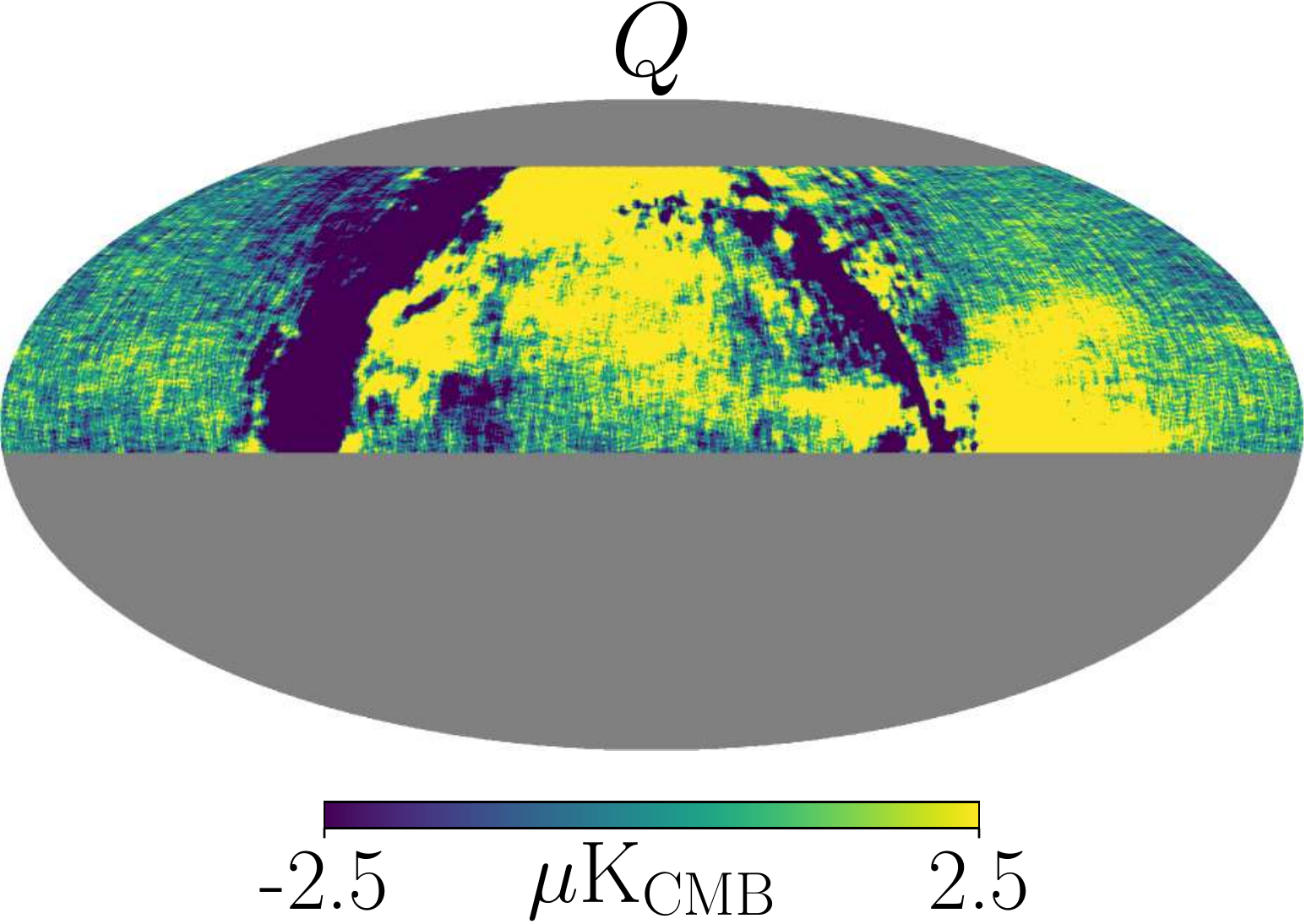}{0.35\textwidth}
            {(a) Frequency map at 145 GHz, $Q$}
            \fig{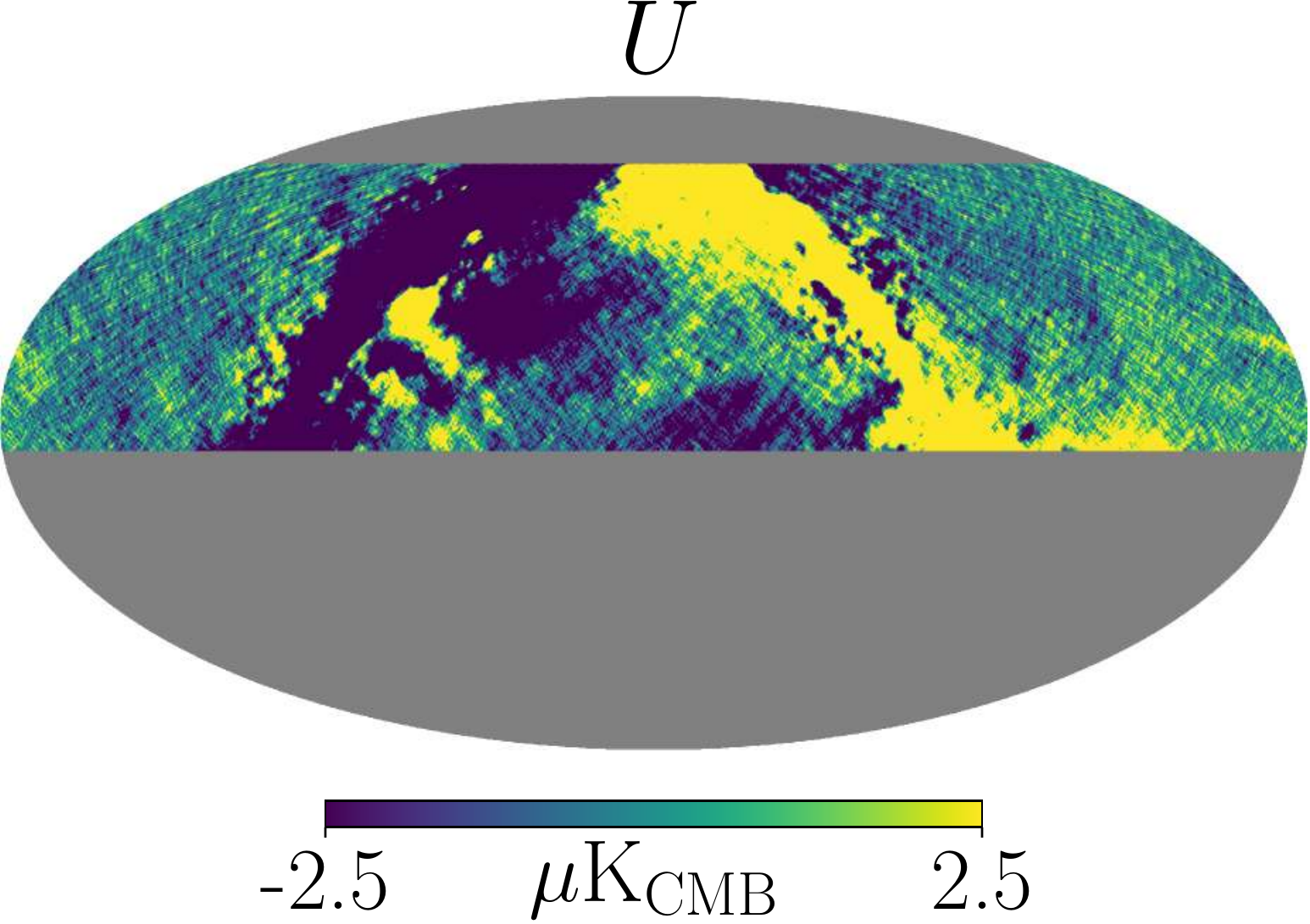}{0.35\textwidth}
            {(b) Frequency map at 145 GHz, $U$}
        }
        \gridline{
            \fig{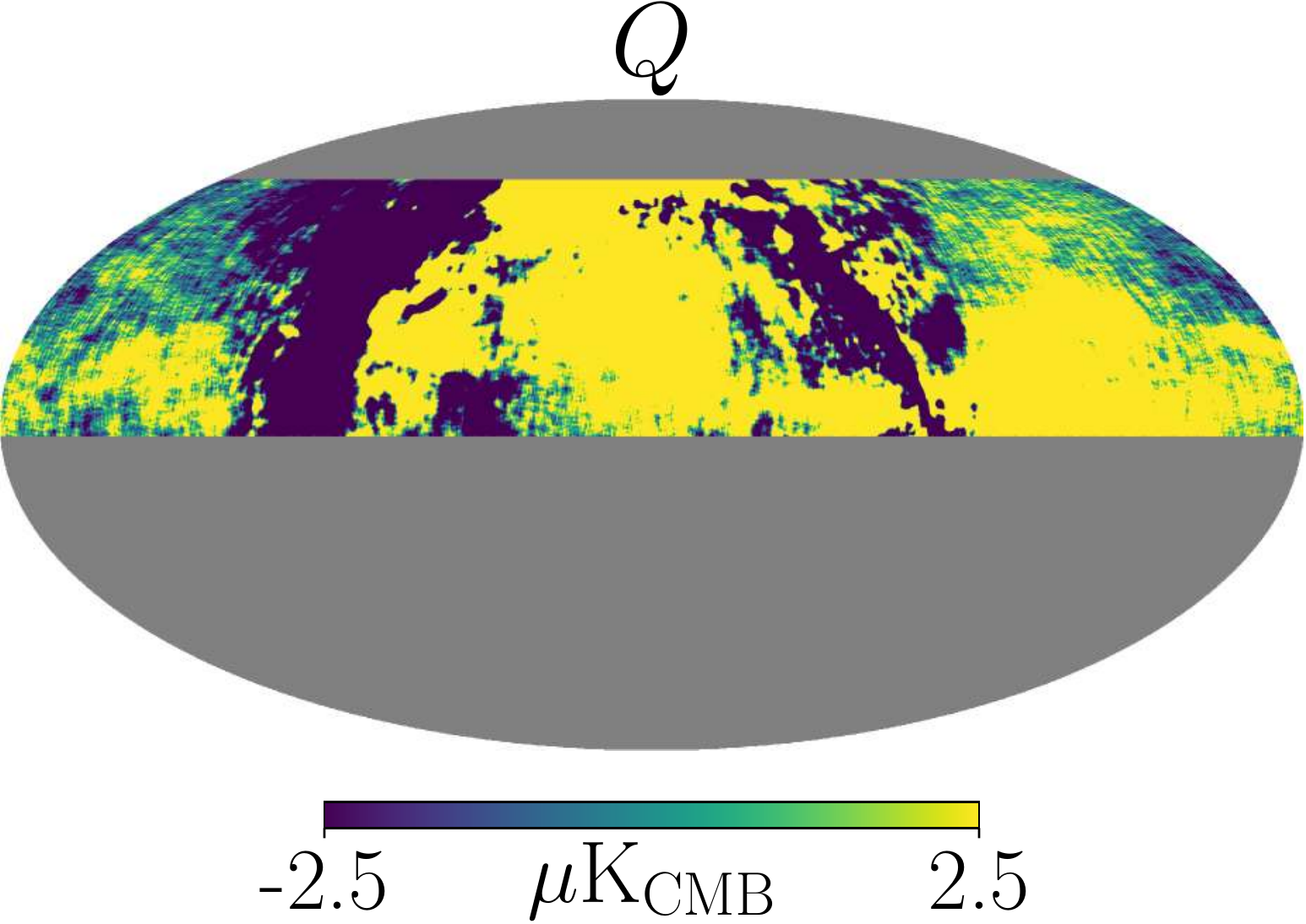}{0.35\textwidth}
            {(c) Frequency map at 220 GHz, $Q$}
            \fig{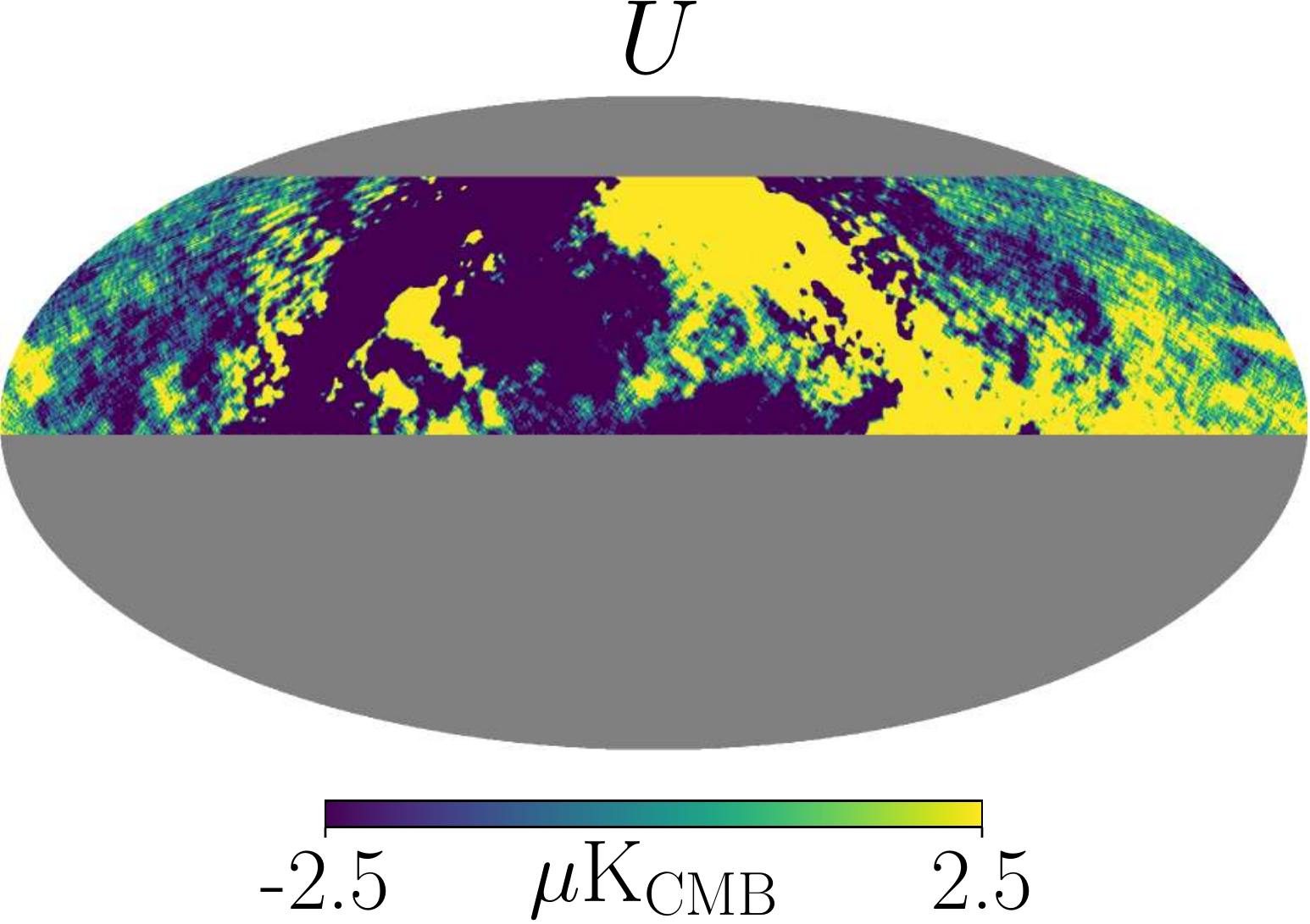}{0.35\textwidth}
            {(d) Frequency map at 220 GHz, $U$}
        }
        \caption{
            Results of map-making for the one-day TOD are displayed. 
            The frequency maps at 145 GHz for the $Q$ and $U$ polarizations 
            are shown in (a) and (b); 
            and the maps at 220 GHz are shown in (c) and (d).
            The equatorial coordinate system is used. 
            The gray region is not observable by the GroundBIRD telescope.
            Here, only CMB and foreground components are included.
        }
        \label{fig:madam_map}
    \end{figure*}
    
The sky maps are constructed from the TOD by the map-making process. 
The $1/f$ component of the TOD needs to be removed before map-making
otherwise ring patterns from the combination of the instrumental variations, change 
in the atmospheric condition, and the scan strategy will remain in the maps.
The process of removing $1/f$ noise is 
referred to as destriping. For destriping and map-making, 
we use MADAM \citep{Keihanen2005, Keihanen2010}. 

Figure \ref{fig:madam_map} shows the result of map-making. 
Since the GroundBIRD telescope observes the same sky area every day, 
we use one-day TOD for map-making to economize computing resources and time.
Accordingly, the map-making residuals and the various types of noise are 
suppressed by assuming three-year observations.
Here, we assume that the map-making residuals are approximately
scaled with the inverse square root of the observation time. 
This assumption was confirmed for our scanning scheme 
by simulating the $1/f$ residuals of the 100-day observations 
and comparing them with the scaled 1-day residuals.
We perform the map-makings separately for each of six modules for 145 GHz and one module for 220 GHz.
The maps for 145 GHz are combined by averaging the individual maps of six modules, 
weighted by the observation time on each sky pixel.
The resulting maps are also pixelized with $N_\textrm{side}=1024$. 
There are ill-defined pixels at the edges of the observed regions 
due to a lack of polarization angle information at such pixels.
To compute the $T$, $Q$, and $U$ from TOD
we need at least two observations at different parallactic angles
to get the correct polarization angle. 
The ill-defined pixels have antenna directions 
that are not distributed over a wide range of angles 
and have too great an uncertainty to obtain correct polarization components. 
We exclude such pixels from our analysis,
and this affects pixels within $0.7\arcdeg$ from the map edge.

After the destriping, there are tiny residuals from the $1/f$ noise. 
These residuals leave unwanted non-zero ring patterns originating 
from the scanning pattern on the maps.
The residuals from the CMB and foregrounds are obtained 
by making the differences between the component maps 
from the map-making and the TOD template maps.
For the $1/f$ noise, the residual is made from the TOD with only $1/f$ noise included. 
The map-making residuals are dominated by 
the $1/f$ residuals. 
The root mean square (RMS) values of polarization intensity maps of the CMB and foreground residuals are 0.03 and 0.07 $\mu \mathrm{K~arcmin}$, respectively,
where the RMS values of $1/f$ residuals are 5.2 $\mu \mathrm{K~arcmin}$ at 145 GHz and 55 $\mu \mathrm{K~arcmin}$ at 220 GHz.
We also check potential bias due to the residuals from the map-making.

\subsection{Sky coverage and mask for the analysis} 

    \begin{figure*} 
        \gridline{
            \fig{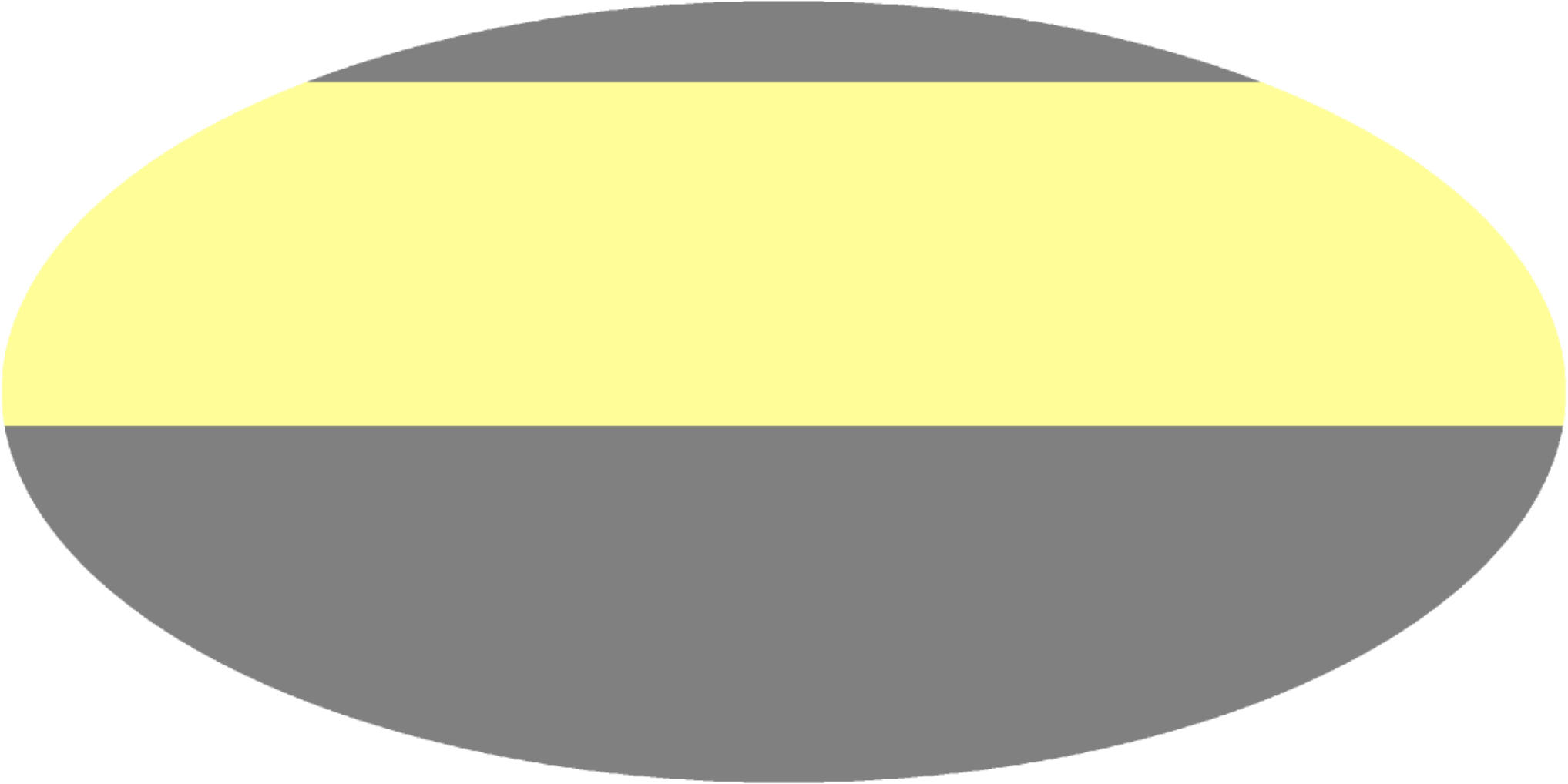}{0.35\textwidth}{(a) GroundBIRD sky coverage}
            \fig{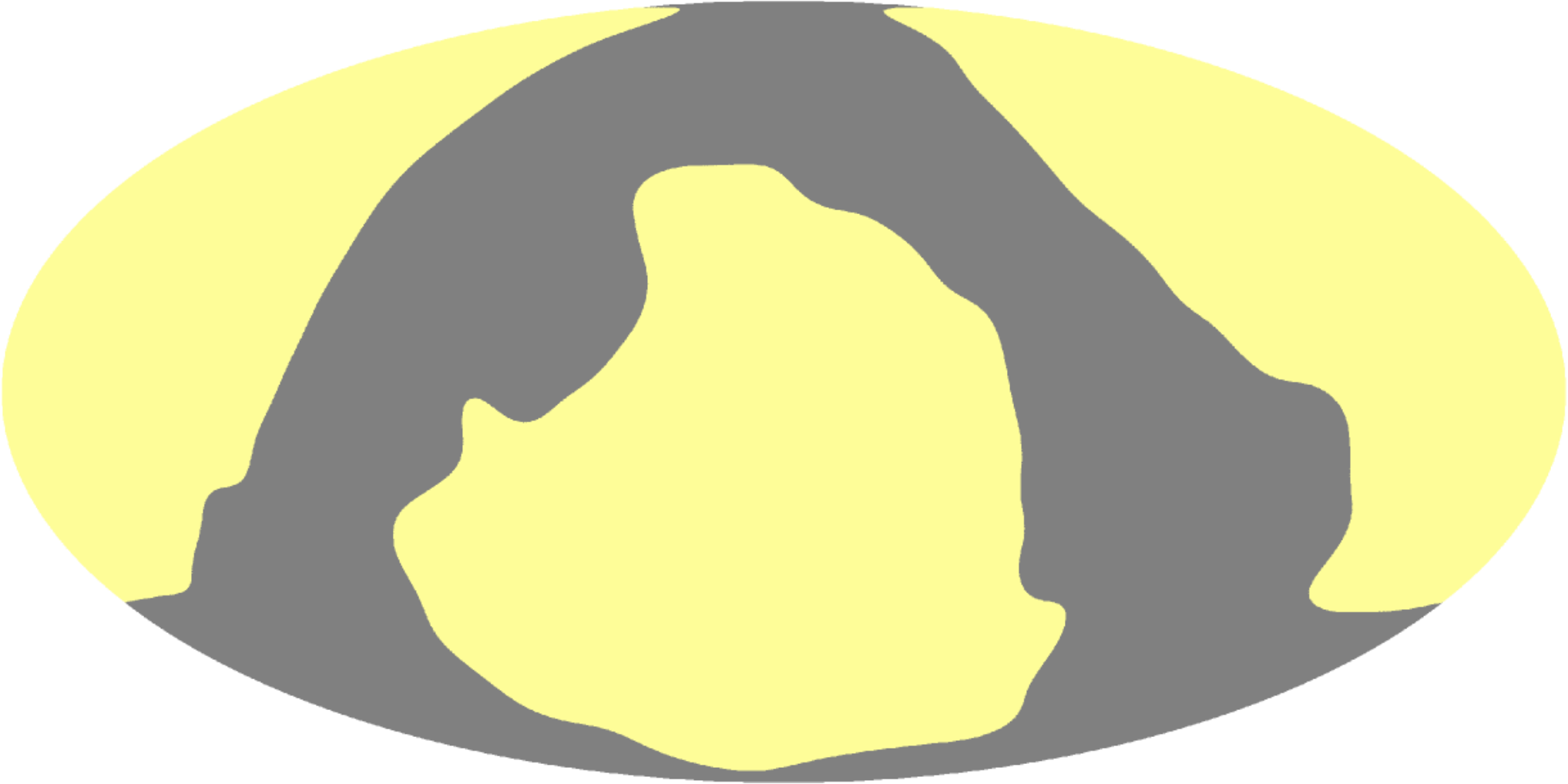}{0.35\textwidth}{(b) Galactic plane mask}
        }
        \caption{
        The GroundBIRD sky coverage and the Galactic plane mask are shown in (a) and (b), respectively. 
        The equatorial coordinate system is used for visualization. 
        $N_\textrm{side} = 1024$ is assumed. 
        The gray regions are masked.}
        \label{fig:mask}
    \end{figure*}

The observing strategy sets the sky coverage of GroundBIRD.
Our simulation gives 
the sky coverages of 0.537 for 145 GHz and 0.462 for 220 GHz. 
Note that the sky coverage for a given module depends on the location of 
the module on the focal plane because our telescope rotates at a fixed elevation.
We use the 220 GHz sky coverage mask for the analysis,  
which masks the outside of a $50 \arcdeg$ wide ring area, as shown in Fig. \ref{fig:mask} (a). 
In Table \ref{tab:gbinfo}, we summarize the number of detectors, the sky coverage, and the pixel noise levels of the GroundBIRD telescope.

\begin{table}[]
    \caption{A summary of the number of detectors, the sky coverage, the mean observation time per detector, and the pixel noise level for the polarization maps of the GroundBIRD telescope is shown. An observation efficiency of 70\% is assumed.}
    \centering
    \begin{tabular}{l|r|r}
        \hline 
         & 145 GHz & 220 GHz \\
        \hline 
        Number of detectors & 138 & 23 \\
        Sky coverage & 0.537 & 0.462 \\
        Mean observation time ($\mathrm{s/arcmin^2}$) & 0.83 & 0.97  \\
        NET ($\mathrm{\mu K ~\sqrt{s}}$) & 820 & 2600 \\
        Pixel noise level ($\mathrm{\mu K ~arcmin}$) & 110 & 780  \\
        \hline 
    \end{tabular}
    \label{tab:gbinfo}
\end{table}

To mask the Galactic plane, we apply the Galactic mask for 
polarization\footnote{{\tt COM\_Mask\_Likelihood-polarization-143\_2048\_R2.00.fits} 
from the Planck Legacy Archive \url{https://pla.esac.esa.int/pla/}}
with a coverage of 52\% used in the Planck analysis \citep{Planck2015xi}, as shown in Fig. \ref{fig:mask} (b). 
The combination of the GroundBIRD sky coverage and the Galactic mask results in coverage of 22.8\% of the full sky.

\section{Foreground cleaning} 
\label{sec:foregroundcleaning}

We use the internal linear combination (ILC) method to remove the foreground components \citep{Eriksen2004}.
A frequency map $\mathbf{m}^{\nu}$ from observations at frequency $\nu$ contains CMB ($\mathbf{m}_{\mathrm{CMB}}$), 
foreground ($\mathbf{m}^\nu_{\mathrm{\{sync,dust\}}}$),
and noise ($\mathbf{m}^\nu_{\mathrm{noise}}$) components. 
The map may be expressed in the form 
\begin{equation}
    \label{eq:freqmap}
    \mathbf{m}^{\nu} = \mathbf{m}_{\mathrm{CMB}} 
                     + \mathbf{m}^\nu_{\mathrm{sync}} 
                     + \mathbf{m}^\nu_{\mathrm{dust}} 
                     + \mathbf{m}^\nu_{\mathrm{noise}},
\end{equation}
where the units of all the maps are $\mu\mathrm{K_{CMB}}$.
The foreground components have different frequency dependencies, whereas the CMB component is constant over frequencies. 
 
We find 
the ILC coefficients for 
the linear combination of observed frequency maps
that minimizes the variance of the combined map ($\mathbf{m}_\mathrm{combined}$), as 
\begin{equation}
    \mathbf{m}_\mathrm{combined} = \sum_{\nu} c_{\nu} \mathbf{m}^{\nu},
\end{equation}
where $c_\nu$ is the ILC coefficient given the frequency $\nu$.
The linear combination of frequency maps
minimizes the foreground components, assuming the sum of the coefficients is 1, or $\sum\limits_\nu c_\nu = 1$.

\section{Likelihood fits} \label{sec:likelihood}

We use the exact likelihood function in real space (or also called the pixel-based likelihood) defined as
    \begin{eqnarray}
        \mathcal{L} (C_\ell) = \frac{1}{|2\pi \mathbf{M}|^{1/2}}
        \exp \left(
        -\frac{1}{2} \mathbf{m}^T \mathbf{M}^{-1} \mathbf{m} 
        \right),
    \end{eqnarray}
where $\mathbf{m}$ is a vector of the temperature map $T$ and and two Stokes parameters $Q$ and $U$. 
The pixel covariance matrix $\mathbf{M}$ is defined by \cite{tegmark2001}
    \begin{equation}
        \mathbf{M} = 
        \left(
        \begin{array}{ccc}
            \langle T_i T_j \rangle & \langle T_i Q_j \rangle & \langle T_i U_j \rangle \\
            \langle T_i Q_j \rangle & \langle Q_i Q_j \rangle & \langle Q_i U_j \rangle \\ 
            \langle T_i U_j \rangle & \langle Q_i U_j \rangle & \langle U_i U_j \rangle 
        \end{array}
        \right),
    \end{equation}
where indices $i$ and $j$ run over pixels. Each $T_i$ is a linear combination of normal spherical 
harmonics while $Q_i$ and $U_i$ are linear combinations of spin-2 spherical harmonics \citep{tegmark2001}. 
The operation of $\langle\ \rangle$ is the average over our ensemble. The ensemble is a set of 
possible realizations of $T$, $Q$, and $U$ maps when the CMB maps have a multivariate Gaussian distribution. 
Since only the polarization maps are used as input data in our analysis, the covariances for polarizations 
$QQ$, $QU$, $UQ$, and $UU$ are used.
We compute the covariance matrix analytically using the formulae in \citet{Zaldarriaga1998}.
The theoretical angular spectra obtained by {\tt CAMB} are used in this calculation.

Note that the use of the real-space likelihood approach ensures mathematical rigor but is extremely
costly from the computational point of view when $N_\textrm{side}$ becomes large, especially for the inversion of a large covariance matrix. 
For that reason, the real-space likelihood is more feasible for studying
large angular scales, which is the case for GroundBIRD. 
In an alternative approach, the likelihood in the harmonic space produces 
unwanted $E\rightarrow B$ mixing for partial sky maps. 
This mixing can be removed by modifying the likelihood \citep{chon2004, smith2007}. 
We choose to use the exact real-space likelihood to study directly 
the sensitivity of the optical depth to reionization with GroundBIRD 
in order to simplify our study. For that reason, we degrade all the input
maps to $N_\mathrm{side}=8$, which corresponds to the maximum multipole moment, 
$\ell_\mathrm{max}=23$ and an angular resolution of $7.3 \arcdeg$. 
We, therefore, need to compute the covariance matrix for $\ell \le 23$. 

The evaluations of the inversion and determinant of covariance matrices are necessary for the likelihood computation.
However, the analytically calculated covariance matrices are singular in most cases, 
and we cannot evaluate the inversion matrices and the determinants correctly.
To bypass this problem, we apply a regularization method \citep{Ledoit2004}. Adding a 
regularization matrix to the covariance matrices allows us to calculate the inversion 
and the determinant successfully. The regularization matrix is created by multiplying
an identity matrix by a small number. The multiplication factor is determined so that 
the covariance matrix is regularized without significantly changing the original matrix.

The calculation of theoretical angular spectra with {\tt CAMB} creates a time-consuming  bottleneck at each iteration of the pixel-based likelihood fit. 
We use a linear interpolation algorithm to improve the computation speed, which is motivated by \cite{Watts2018}. 
A set of the pre-computed angular power spectra is evaluated by sweeping the parameter space with {\tt CAMB}. 
The angular power spectra for specific parameters are obtained by linear interpolation between the pre-computed spectra.
This method is proper for less than four parameters because the time consumption and the 
storage space for the pre-computed data are exponentially proportional to the number of parameters.
We make the pre-computed data set for $\tau$ by sampling 100 points.
This method improves the calculation speed by a factor of 100 with only a negligible difference from the result obtained by {\tt CAMB}.

We perform an ensemble test for 10\,000 realizations of the CMB and noise
to examine possible fit bias and estimate the expected uncertainty on $\tau$.
For each realization, we generate a set of input maps. 
If we follow the whole simulation procedure for every realization, the TOD simulation and 
the map-making will take several years and require 
inaccessible computing resources. 
We therefore simplify the data simulations by skipping the time-consuming steps.
For the ensemble test input maps, we make CMB and noise maps for every realization. 
We make the CMB maps directly by {\tt SYNFAST} as we did for the TOD template.
We also use {\tt SYNFAST} to make the white noise maps for each frequency with the assumed pixel noise levels.
The foregrounds are constant over the realizations and need to be made only once.
The input map combines the CMB map and the white noise map for the realization, and the fixed foreground map.
Additionally, we include the map-making residuals for the 145 and 220 GHz maps to evaluate the systematic effect of the $1/f$ noise. 
We compare the sum of each component map generated independently with the map
from the single map-making with the TOD, including all the components.
We find that the difference is negligible compared to any component or 
map-making residuals. Thus, we can form a frequency map by combining
independently generated component maps.
With the simulated maps, we remove the foreground components with the ILC method and 
then perform the maximum likelihood fit.
The fit parameters are $\tau$ and the total noise level $\sigma_p$. 
The total noise level is the pixel noise level of the combined polarization map 
after foreground cleaning. 

We verify the above technique using simulated frequency maps for the Planck frequencies. 
We simulate the maps with our simulation pipeline for all of the Planck frequency bands (30, 44, 70, 100, 145, 217, and 353 GHz). 
The white noises are generated with the noise levels obtained from the published Planck frequency maps. 
The uncertainty of $\tau$ for the simulated Planck data is found as $\sigma_{\tau} = 0.007$, 
which is compatible with the published value \citep{planck2018vi}.
The small discrepancy between our value and the published value 
arises because the analysis methods are different 
and our simulated maps are based on the aforementioned assumptions.


\section{Results} \label{sec:result}

First, we show the result with the GroundBIRD simulations only and then
the result with three 
frequency maps including a synchrotron template. We also show the result with the GroundBIRD and QUIJOTE simulations here.
The cases without foregrounds or map-making residuals are summarized in Appendix \ref{appendix:results}.

\subsection{GroundBIRD only}

    \begin{figure}
        \includegraphics[width=0.47\textwidth]{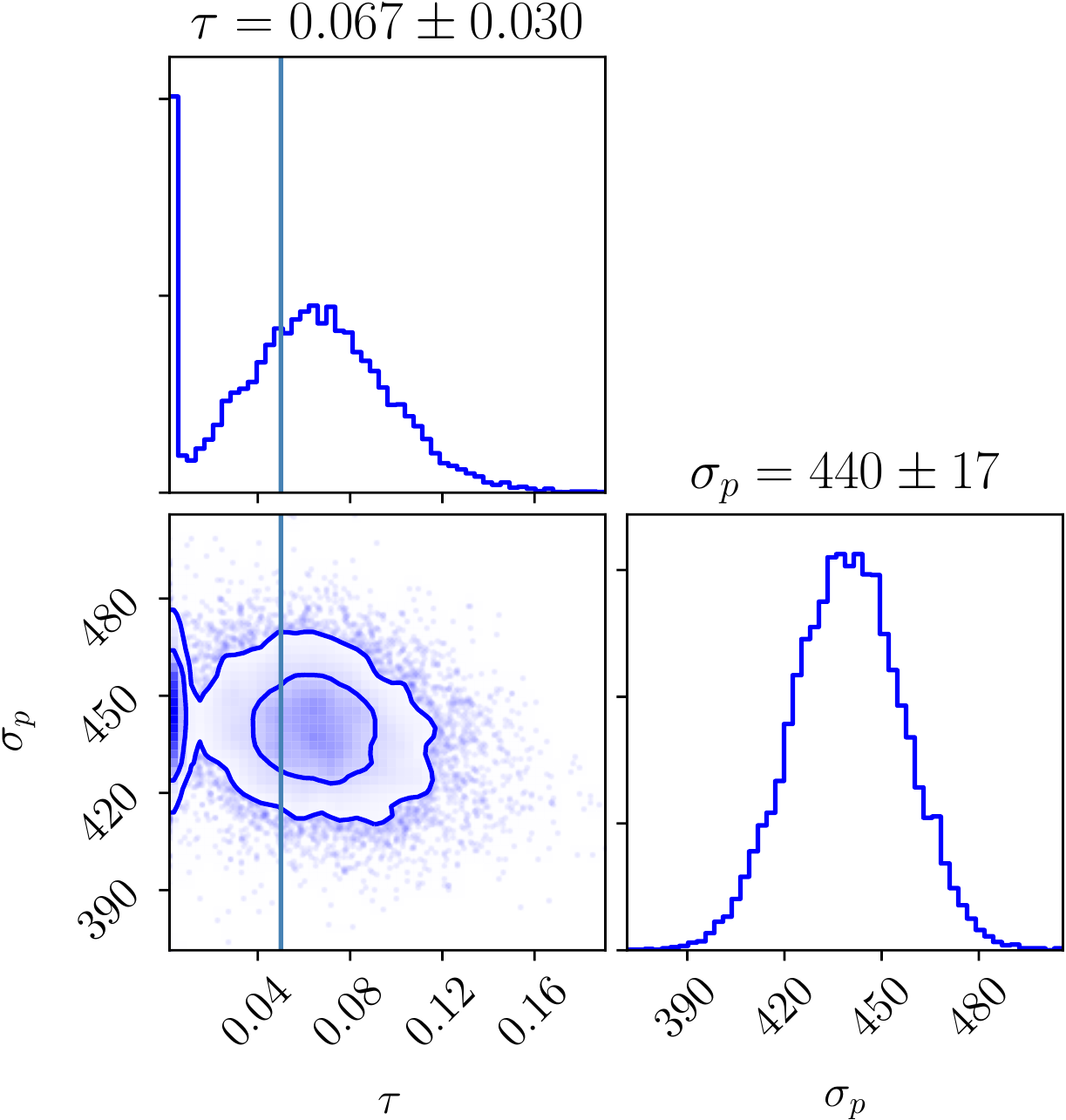}
        \caption{The result of the ensemble test with the two-frequency ILC is shown. 
        The simulated GroundBIRD frequency maps are used. 
        The contours for $1\sigma$ and $2\sigma$ are drawn, 
        and the vertical solid lines in the $\tau$ histogram 
        and contours display the input value of $\tau$. 
        We estimate $\tau$ from the mean and standard deviation of its distribution excluding the peak at the lower boundary.
        The estimated $\tau$ is 0.067$\pm$0.030 with the 
        total noise level estimate of 440$\pm$17 $\mu \mathrm{K~arcmin}$. 
        }
        \label{fig:contour_ilc2}
    \end{figure}

We first fit the GroundBIRD maps. The ensemble test is performed with 10\,000 samples.
For each realization, we remove the foreground first and then estimate $\tau$ from a pixel-based likelihood fit.

The ILC with the two GroundBIRD maps leaves 
a large foreground residual mainly from the synchrotron component. 
This residual prevents the correct estimation of $\tau$. 
To solve this problem, we add a mask to exclude the sky region dominated by large synchrotron residuals 
from the input map for the likelihood fit. This mask is obtained by applying a threshold to the 
foreground cleaned map. This extra mask excludes an additional $5\%$ of the sky 
and enables estimation of the unbiased $\tau$.
After applying this mask, the ILC coefficients for 145 and 220 GHz are 
$c_{145}$=1.470$\pm$0.003 and $c_{220}$=$-$0.470$\pm$0.003,
respectively. 

The likelihood fit result for the ensemble is shown in Fig.\ \ref{fig:contour_ilc2}, 
where a contour of $\tau$ versus $\sigma_p$ and two histograms of them are shown. 
The estimated $\tau$ is found to be 0.067$\pm$0.030 when the input $\tau$ is 0.05. 
The noise level of the foreground cleaned map is found to be 
440$\pm$17 $\mu \mathrm{K~arcmin}$, 
to which the white noise and map-making residuals are contributing.

A small peak near the lower boundary of the $\tau$ histogram
in Fig. \ref{fig:contour_ilc2} originates from our high noise level.
Fluctuation in an ensemble, which simulates the statistical uncertainty of $\tau$, 
sometimes pushes $\tau$ down to its lower limit that is 0, making such a peak.
One way to avoid it is to allow the negative value of $\tau$
in the likelihood. However, since it is not only unclear how to
correctly construct the pixel likelihood with the unphysical negative
$\tau$ values from the {\tt CAMB}, but is also not of the main issue of
this paper, we limit $\tau$ to non-negative values.
We estimate $\tau$ from the mean and standard deviation of its distribution, excluding this peak.

Even after applying the extra mask for the synchrotron, 
the dominant foreground residual 
at this stage originates from the synchrotron foreground. 
We find that the non-zero foreground residuals introduce a bias in $\tau$, which 
tends to be overestimated by around 10\% but is still within $1\sigma$ of the statistical uncertainty. 

To remove the synchrotron foreground further, 
we add the simulated QUIJOTE 30 GHz map as a synchrotron template.
The ILC coefficients for the three-frequency case are 
$c_{145}$=1.485$\pm$0.007, 
$c_{220}$=$-$0.468$\pm$0.003, 
and $c_{30}$=$-$0.017$\pm$0.006.
The ILC coefficient for 30 GHz ($c_{30}$) is suppressed 
because the synchrotron foreground dominates over other components at this frequency.
Therefore, the synchrotron template removes the synchrotron 
foreground without significantly changing the map.
The estimated $\tau$ is 0.059$\pm$0.029
and the noise level is found to be 
440$\pm$17 $\mu \mathrm{K~arcmin}$. 

Note that
the map-making residuals become practically an additional white noise source
because their ring patterns are diluted as the maps are degraded to 
$N_\mathrm{side} = 8$.
We find that the existence of the map-making residuals 
increases the noise level effectively by 11\%
and thus the uncertainty of the $\tau$ is increased by around 10\%. 
We discuss the results with and without map-making residuals in Appendix \ref{appendix:results}.

\subsection{GroundBIRD and QUIJOTE}

    \begin{figure}
        \includegraphics[width=0.47\textwidth]{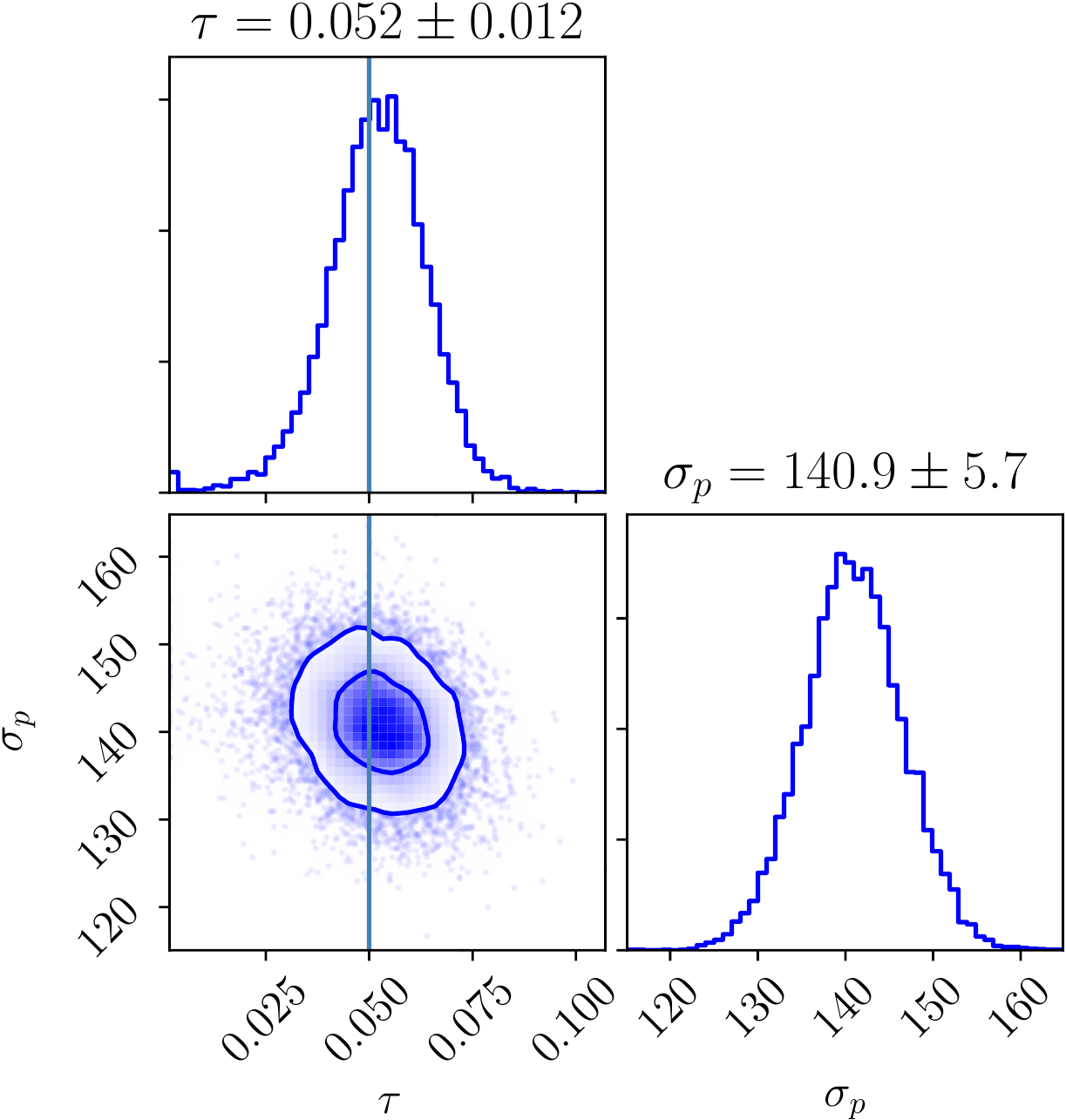}
        \caption{The result of the ensemble test with the eight-frequency ILC is shown. 
        The simulated GroundBIRD maps for \{145, 220\} GHz 
        and the simulated QUIJOTE maps for \{11, 13, 17, 
        19, 30, 40\} GHz are used. The contours for 
        $1\sigma$ and $2\sigma$ are drawn, 
        and the vertical solid lines in the $\tau$ histogram 
        and the contour plot display the input value of $\tau$. 
        The estimated $\tau$ is 
        0.052$\pm$0.012.
        }
        \label{fig:contour_ilc8}
    \end{figure}

We estimate $\tau$ again by including the simulated QUIJOTE frequency maps 
at 11, 13, 17, 19, 30, and 40 GHz in our analysis. 
We generate the foreground and noise maps for QUIJOTE in a similar way as for the GroundBIRD maps. 
Here the $1/f$ noise component is not included in the QUIJOTE maps and is a potential limitation of our analysis. We discuss this point in the following section.
The pixel noise levels of the polarization maps and the resulting ILC coefficients are summarized in Table \ref{tab:ilc8}.
In this case, the map-making residuals are included only in the GroundBIRD maps,
and the ILC coefficients of the GroundBIRD maps are relatively smaller than in previous cases. 
The effect of the map-making residuals, therefore, becomes small.

\begin{table}[ht!]
    \caption{Noise levels of polarization maps and ILC coefficients for the eight-frequency result are summarized. The constraint of $\sum_{\nu} c_{\nu} = 1$ is numerically incorrect due to the rounding off of fit results.}
    \centering
    \begin{tabular}{l  r | r |  r@{$\pm$}l}
        \hline 
        &Frequency & Noise level & \multicolumn2c{ILC coefficients} \\
        &(GHz) & ($\mu\mathrm{K~arcmin}$) & \multicolumn{2}{c}{}  \\
        \hline 
        & 11 & 3600 & $-$0.011 & 0.001\\
        & 13 & 3600 & $-$0.009 & 0.002\\
        & 17 & 5100 & $-$0.003 & 0.002\\
        & 19 & 5100 & $-$0.002 & 0.002\\
        & 30 & 160  & $-$0.14  & 0.04\\
        & 40 & 91   &    1.12  & 0.04\\
        &145 & 110  &    0.13  & 0.03\\
        &220 & 780  & $-$0.077 & 0.008\\
        \hline 
    \end{tabular}
    \label{tab:ilc8}
\end{table}

The result of the eight-frequency case is shown in Fig. \ref{fig:contour_ilc8}.
The estimated $\tau$ is found to be 0.052$\pm$0.012 with the total noise level of 141$\pm$6 $\mu\mathrm{K~arcmin}$.
The noise level is reduced by 70\% compared to the three-frequency result 
because the contribution of the 220 GHz map with a high noise level is suppressed dominantly by adding 30 and 40 GHz maps. 
Accordingly, the sensitivity on the $\tau$ is improved 
compared to the three-frequency case. 

\subsection{Limitations of current analysis}

We include several sources of systematic errors for realistic simulations:
atmospheric noise, detector calibration, and map-making residuals. 
However, there are potential sources of systematic error that are not included in this study. 

We assume that our fast-scanning technique, 
in combination with the sparse wire-grid calibration 
removes most of the atmospheric noise. 
Therefore, we do not consider the effect of frequency-dependent atmospheric fluctuations  \citep{Lay2000} in this work.
The systematic effects from the optical system are incomplete. 
For example, the uncertainty of the polarization angle and imperfect focusing due to 
the aberration of the reflectors are not included in our analysis. 
The detailed study for the sidelobes and spillover patterns is also in progress.

In our TOD simulation, we do not consider the systematic effects due to data cuts,
potential pointing offsets and detector gain variations.
These should be addressed when analyzing the real observations.

As is mentioned earlier, the foreground and noise maps for QUIJOTE do not contain the $1/f$ noise, since the behavior of $1/f$ noise for QUIJOTE is not available to us at this stage. Once this information is available, we plan to include it in our analysis.
We will address all the issues discussed here in a future publication.


\section{Conclusions} \label{sec:conclusions} 

We have performed estimates of the optical depth to reionization ($\tau$)
with simulated data for the GroundBIRD experiment.
We started the simulation from TOD with the inclusion of $1/f$ noise in our simulation.
The input maps cover 45\% of the sky (22.8\% after masking the Galactic plane).
These maps included the CMB realization, foregrounds, expected noise, and map-making 
residuals. We assumed a three-year observation with 161 detectors in total.
We cleaned the polarization foregrounds using the ILC method, which requires at least three 
frequency maps to remove two foreground components; namely, synchrotron and dust foregrounds. 
We also added the simulated QUIJOTE maps for better foreground cleaning. 
To estimate the $\tau$ values, we used our pixel-based likelihood.

The estimated $\tau$ is 0.067$\pm$0.030 with GroundBIRD maps only and 0.059$\pm$0.029 with a 
synchrotron template included, where the input value is 0.05.
Adding QUIJOTE frequency maps to the GroundBIRD maps further removed the foregrounds and reduced the noise level. 
The estimated $\tau$ with eight-frequency (11, 13, 17, 19, 30, 40, 145, and 220 GHz) ILC is found to be 0.052$\pm$0.012.
The incomplete foreground cleaning tends to introduce additional bias into the $\tau$ estimate.
The uncertainty in the
$\tau$ estimate is constrained mainly by the pixel noise level of the foreground cleaned map. The map-making residuals increase the noise level and degrade the sensitivity to $\tau$. 
In our case, $\tau$ sensitivity is degraded by about 10\%, which depends on the observing strategy and detector set-up.

\acknowledgments
The authors would like to thank the Instituto de Astrof\'isica de Canarias 
for their support and hospitality 
during the installation of the GroundBIRD telescope at Teide Observatory.
This work was partially supported 
by the National Research Foundation of Korea (NRF) grant funded
by the Korea government (MSIT) (No. NRF-2017R1A2B3001968) and Korea University Grant.
This work was also supported 
by the JSPS Grant Number JP15H05743, JP18H05539, and Heiwa Nakajima Foundation.
The computing network was partially supported by KISTI/KREONET.
Some of the results in this paper have been derived using the {\tt HEALPix} package.

\software{
    {\tt astropy} \citep{astropy:2013,astropy:2018},
    {\tt CAMB} \citep{Lewis:1999bs},
    {\tt corner} \citep{corner}
    {\tt HEALPix} \citep{healpix},
    {\tt healpy} \citep{healpy},
    {\tt iminuit} \citep{iminuit},  
    {\tt python} \citep{python}, 
    {\tt PySM} \citep{PySM}
}

\appendix

\section{Estimation of noise equivalent temperature of GroundBIRD}
\label{appendix:NET}

We estimate our noise level by
considering the KID response, the optical efficiency, 
and the atmospheric emission. 
The noise equivalent power (NEP) is given by \citep{Zmuidzinas2012,Flanigan2016}

\begin{equation}
    \mathrm{NEP}^2 = \frac{1}{\eta_\mathrm{opt}} \int d\nu \left[ 
    2h\nu P_\mathrm{rad}(\nu)
        \left\{
            1+\eta_\mathrm{opt}\eta_\mathrm{em}(\nu)\bar{n}_\mathrm{bb} F(\nu)
        \right\}
        +\frac{4\Delta}{\eta_\mathrm{pb}} P_\mathrm{rad}(\nu) 
    \right]
\end{equation}
where $h$ is the Planck constant, 
$\nu$ is the frequency of the incoming radiation,
$P_\mathrm{rad}(\nu)$ is the radiation power density in $\mathrm{pW/\sqrt{Hz}}$,
reaching the telescope window at a given elevation angle, and $\eta_{\mathrm{em}}(\nu)$ is the sky emissivity that depends on the frequency.
To compute $P_\mathrm{rad}(\nu)$ and $\eta_\mathrm{em}(\nu)$,
we use the atmospheric antenna temperature for 
airmass of $1/\sin(70^{\arcdeg})$ derived from the  atmospheric transmission at microwave model \citep{Pardo_2001_atm}, 
which is evaluated assuming average conditions at the Teide observatory (pressure of 770.0 mbar and precipitable water vapor (PWV) of 3.5 mm \citep{pwv_tenerife}).
For the actual modeling, we assume the PWV value to be 4.0 mm.

The $\eta_{\mathrm{opt}}$ is the net optical efficiency of our optical system, 
$\bar{n}_{\mathrm{bb}}$ is the mean photon number emitted from blackbody,
$\eta_{\mathrm{pb}}$ is the Cooper-pair breaking efficiency,
and $F(\nu)$ is the total transmissivity 
of the filters that are used in GroundBIRD optics.
The $\Delta$ is the gap energy of the superconducting material which is given by $1.76 k_B T_c$ where $k_B$ is the Boltzmann constant and $T_c$ is the superconducting transition temperature ($T_c=1.28\ \mathrm{K}$ for the aluminum thin film \citep{Kutsuma2020}).
The NEP calculation assumes the atmosphere to be a blackbody source.
The parameters and their values used for this estimation are summarized in Table \ref{tab:etas}.
The frequency-dependent parameters are shown by nominal values in our frequency bands, and the radiation power ($\mathcal{P}_\mathrm{rad}$) is integrated over our frequency band with a width of 60 GHz, for both 145 and 220 GHz. 
The estimated NEP are $160\ \mathrm{aW\sqrt{s}}$ and $290\ \mathrm{aW \sqrt{s}}$ for 145 GHz and 220 GHz, respectively. The corresponding NET values in CMB temperature are $820\ \mu \mathrm{K~\sqrt{s}}$ and $2600\ \mu \mathrm{K~\sqrt{s}}$ for 145 GHz and 220 GHz, respectively.

    \begin{table}[h]
        \centering
        \begin{tabular}{ c  c  c }
            \hline 
            Parameter & 145 GHz & 220 GHz \\
            \hline 
            $\mathcal{P}_\mathrm{rad}$ [$\mathrm{pW}$] 
                                           & 13   & 28 \\
            $F(\nu)$                       & 0.5  &  0.5 \\
            $\eta_\mathrm{em}(\nu)$        & 0.15 &  0.28 \\
            $\eta_\mathrm{opt}$            & 0.38 &  0.30 \\
            $\eta_\mathrm{pb}$             & 0.4  &  0.4 \\
            $\bar{n}_\mathrm{bb}$          & 38.7 & 25.4 \\
            \hline 
        \end{tabular}
        \caption{The parameters and their nominal values at two frequency bands used for the estimation of NEP of the GroundBIRD detectors are summarized.}
        \label{tab:etas}
    \end{table}

\section{Likelihood fit results for other set-ups}
\label{appendix:results}
 
In this section, we summarize the results of ensemble tests for several foreground set-ups.
The estimates of $\tau$ and total noise level $\sigma_p$ are summarized in Table \ref{tab:tauestimations}. 
The simplest case is the ``No foregrounds'' case, where it is assumed that the foreground is cleaned completely. The 
input map contains CMB and noise only, and the same noise level as the ``Two-frequency ILC'' case is assumed. 
The estimated $\tau$ is 0.060$\pm$0.029 when the input value is 0.05, with an estimate of the white 
noise level of 397$\pm$17 $\mu\mathrm{K~arcmin}$.
This uncertainty is 7\% greater
than the estimate by Fisher matrix, $\sigma_{\tau} $= 0.027, with the same noise input.
The estimated $\tau$ has a large bias ($\sim 0.01$) from the input value 
but it is still within the uncertainty. 
We test this case with smaller noise levels and find that
a large noise tends to make a positive bias in $\tau$ estimation.

The cases with ILC are tested in two ways: with and without map-making residuals. 
For the two-frequency case, the estimated $\tau$ is 0.066$\pm$0.028, and $\sigma_p$ is 395$\pm$16 $ \mu \mathrm{K~arcmin}$
without the map-making residuals. With the map-making residuals, the estimated $\tau$
 is 0.067$\pm$0.030, and $\sigma_p$ is 440$\pm$17 $\mu \mathrm{K~arcmin}$. 
The map-making residuals increase the noise level by 10\% and also increase the $\tau$ uncertainty by around 10\%. 
As discussed in section 6.1, incomplete foreground cleaning with two frequency maps tends to create an additional bias of around 10\% in $\tau$ estimate.

In the three-frequency cases, 
we expect the synchrotron to be removed effectively by adding a 30 GHz map. 
Without the map-making residuals, 
the estimated $\tau$ is 0.056$\pm$0.026, 
with the estimated noise level of 394$\pm$15 $\mu \mathrm{K~arcmin}$. 
The $\tau$ uncertainty and the white noise level are 
at almost same level as the two-frequency map case.
Adding a synchrotron template map at 30 GHz cleans the synchrotron foreground better, so the estimated $\tau$ is closer to the input value.
With the map-making residuals, 
the estimated $\tau$ is 0.059$\pm$0.029 
which has about 10\% larger uncertainty compared to the value without map-making residuals.
The noise level is almost same with the two-frequency case.

The eight-frequency map case uses all six QUIJOTE frequency maps 
with the GroundBIRD frequency maps. 
The map-making residuals are included only for the GroundBIRD frequency maps 
because we are not doing map-making for the QUIJOTE maps.
The estimated value of $\tau$ is 0.052$\pm$0.012
and noise level is 139$\pm$6 $\mu \mathrm{K~arcmin}$ without 
the map-making residuals and $\tau$ is 0.052$\pm$0.012 
and noise level is 141$\pm$6 $\mu \mathrm{K~arcmin}$ with the map-making residuals. 
Since the noise levels are smaller than the two or three-frequency cases, the bias in $\tau$ is suppressed.
In this case, the map-making residuals slightly increase the noise level, but have a negligible effect on the estimation of $\tau$ 
because the contribution of the GroundBIRD 220 GHz map, which 
has a large white noise level and the map-making residuals, becomes small by adding the QUIJOTE maps.
    
    \begin{table}
        \caption{ 
            The estimates of $\tau$ and the pixel noise level of the combined map $\sigma_p$ for 
different foreground cleaning methods are summarized.
            For the ``Fisher matrix'' case, the same noise level as the two-frequency map case is assumed and
 the $\tau$ uncertainty is estimated from the Fisher matrix approach. Other estimates are from the pixel-based likelihood approach. 
            The ``No foregrounds'' is the case with an input map combining only the CMB and noise. The same noise with 
``Two-frequency ILC'' case is assumed.
            The ``Two-frequency ILC'' is the case performing the foreground cleaning with only two frequency maps of GroundBIRD. 
            For the ``Three-frequency ILC'' case, a 30 GHz map is added to the two-frequency case, as a synchrotron template.
            The``Eight-frequency ILC'' is the case with eight frequency maps, including all the QUIJOTE frequency maps.
            The cases with foregrounds are tested with and without the map-making residuals.} 
        \centering
        \begin{tabular}{l r @{$\pm$} l r @{} l}
            \hline
            Foreground cleaning method & \multicolumn{2}{c}{$\tau$} & \multicolumn{2}{c}{$\sigma_p\ (\mu \mathrm{K~arcmin})$} \\
            \hline 
            Fisher matrix  
            & 0.050 & 0.027 & 395 & \\
            No foregrounds
            & 0.060 & 0.029 & 397 & $\pm$17 \\
            Two-frequency ILC
            & 0.066 & 0.028 & 395 & $\pm$16\\
            Two-frequency ILC with map-making residuals
            & 0.067 & 0.030 & 440 & $\pm$17\\
            Three-frequency ILC 
            & 0.056 & 0.026 & 394 & $\pm$15\\
            Three-frequency ILC with map-making residuals 
            & 0.059 & 0.029 & 440 & $\pm$17 \\
            Eight-frequency ILC
            & 0.052 & 0.012 & 139 & $\pm$~6 \\
            Eight-frequency ILC with map-making residuals 
            & 0.052 & 0.012 & 141 & $\pm$~6 \\
            \hline
        \end{tabular}
        \label{tab:tauestimations}
    \end{table}

\bibliography{GB}
\bibliographystyle{aasjournal}



\end{document}